\documentclass[aps,prd,a4paper,10pt]{revtex4-1}
\usepackage[utf8]{inputenc}
\usepackage[english]{babel}
\usepackage{graphicx}
\usepackage{amsmath, amssymb, amsfonts, amsthm, esint}
\usepackage{slashed}
\usepackage{epsfig}
\usepackage{color}
\usepackage[usenames,dvipsnames,svgnames]{xcolor}
\usepackage[colorlinks=true, urlcolor=DarkBlue, anchorcolor=DarkBlue, citecolor=DarkRed, filecolor=DarkBlue, linkcolor=DarkRed, menucolor=DarkBlue, linktocpage=true]{hyperref}
\usepackage{booktabs}

\newenvironment{eq}{\begin{eqnarray}}{\end{eqnarray}}

\begin{document}

\title{Majoron Dark Matter From a Spontaneous Inverse Seesaw Model}

\author{N.~Rojas}
\email[]{nicolas.rojasro@usm.cl}
\affiliation{Departamento de Física, Universidad Técnica Federico Santa María.\\ Casilla 110-V,Valparaiso, Chile.}
\author{R.~A.~Lineros}
\email[]{roberto.lineros@ucn.cl}
\affiliation{Departamento de Física, Universidad Católica del Norte.\\ Avenida Angamos 0610, Casilla 1280, Antofagasta, Chile.}
\author{F.~Gonzalez-Canales}
\email[]{felixfcoglz@gmail.com}
\affiliation{Fac. de Cs. de la Electrónica, Benemérita Universidad Autónoma de Puebla.\\ Apdo. Postal 542, Puebla, Pue. 72000, México.}

\date{\today}

\begin{abstract}
The generation of neutrino masses by inverse seesaw mechanisms has advantages over other seesaw models since the potential new physics can be produced at the TeV scale. We propose a model that generates the inverse seesaw mechanism via spontaneous breaking of the lepton number, by extending the Standard Model with two scalar singlets and two fermion singlets both charged under lepton number. The model gives rise to a massless Majoron and a massive pseudoscalar which we dub as \emph{massive Majoron}, which corresponds to the Nambu-Goldstone boson of the breaking of lepton number. If the massive Majoron is stable in cosmological time, it might play the role of a suitable Dark Matter candidate. In this scenario, we examine the model with a massive Majoron in the keV range.
In this regime, its decay mode to neutrinos is sensitive to the ratio between the vevs of the new scalars ($\omega$), and it vanishes when $ \omega \simeq \sqrt{2/3}$, which is valid within a large region in the parameter space. On the other hand, the cosmological lifetime for the Dark Matter candidate places constraints on its mass via scalar decays. In addition, simple mechanisms that explain the Dark Matter relic abundance within this context and plausible modifications to the proposed setup are briefly discussed.
\end{abstract}


\maketitle

\section{\label{sec:intro}Introduction}

The success of the Standard Model of particle physics (SM) has been established given 
the experimental observations of its predictions~\cite{Weinberg:1967tq,Higgs:1964pj,Arnison:1983rp,Bagnaia:1983zx,Aad:2012tfa,Chatrchyan:2012xdj},
although it still has to face some theoretical and challenges. One of them is the 
existence of the Dark Matter (DM), which even though is the largest matter component 
in the Universe~\cite{Oort:1932BAN,Rubin:1980zd,Ade:2015xua} (around $85 \%$),
we have few clues about its properties. Another is the neutrino oscillations~\cite{Pontecorvo:1957cp,Maki:1962mu,Ahmad:2001an,Fukuda:1998tw}, 
which is a consequence of the still-not-measured neutrino masses. On this basis, 
from neutrino oscillation experiments we get two of their most striking features: $i)$~the leptonic mixing angles are large (other than the mixing values in the CKM matrix), 
and $ii)$~their mass scale is in the sub-eV range~\cite{Olive:2016xmw,Esteban:2016qun,Capozzi:2016rtj,Forero:2014bxa}.
Of course, one might tackle both problems at once in many ways (see, for instance,~\cite{Lattanzi:2014mia,Hirsch:2013ola,Heurtier:2016iac,deSalas:2016svi}),
however, they are an indication that there must be physics beyond the SM.

A simple SM completion considering Majorana neutrino masses arises after the inclusion
of the dimension-5 Weinberg operator~\cite{Weinberg:1979sa}. To give rise to
a renormalizable theory, we may extend the SM particle content so that the Weinberg 
operator is properly accounted for. This way of reasoning leads to the 
type I~\cite{Minkowski:1977sc,Mohapatra:1979ia}, II~\cite{Schechter:1980gr,Mohapatra:1980yp}, 
and III~\cite{Foot:1988aq} \emph{seesaw mechanisms}. Herein, the fact that
Majorana masses break explicitly the lepton number, an accidental symmetry of the SM,
is heavily exploited. If we choose to break this symmetry via a spontaneous 
symmetry breaking mechanism, we will have a massless pseudoscalar as a physical state,
which is termed as \emph{The Majoron}~\cite{Riazuddin:1981hz, Chikashige:1980ui}.

However, after invoking that global symmetries (for instance, the lepton number) might be broken due to 
Planck scale effects, the Majoron could get a small mass~\cite{Rothstein:1992rh}. 
Jointly with the fact of that the Majoron is neutral and its interactions appear to be 
suppressed, the Majoron is promoted to a \emph{decaying} DM candidate~\cite{Boucenna:2012rc}, 
although with a half-life larger than the age of the Universe~\cite{Berezinsky:1993fm,Lattanzi:2007ux}.
Nevertheless, its main decay channels would be neutrinos, and in a more model-dependent way, 
into photons~\cite{Schechter:1980gr,Berezinsky:1993fm,Aranda:2009yb,Lattanzi:2007ux,Queiroz:2014yna,Wang:2016vfj,Garcia-Cely:2017oco}.

Even though models considering spontaneous breaking of the lepton number symmetry are candidates to solve via Majorons both neutrino and DM puzzles, there is still a lack of a low scale mechanism (i.e, at the TeV scale) explaining them. 
This, given that the three basic types of seesaw mechanism act at scales
as high as $10^{12}$ GeV.
In the literature, there are variations called {\it low scale seesaw}~\cite{Mohapatra:1986aw,Mohapatra:1986bd,Barr:2003nn}
in which new particles with masses of few TeV arise to give rise to the seesaw mechanism. 
These are the so-called \emph{linear} and \emph{inverse} seesaw scenarios~\cite{Mohapatra:1986bd,Abada:2014vea}.

In this work we employ the context of inverse seesaw to embed a completion
of the SM where a rich pseudoscalar sector arises related to the breakdown of the lepton number symmetry.
Therein, a massless Majoron and a Massive Majoron arise.
Thus, the paper is organized as follows: We describe the basics of the inverse seesaw and our model in section~\ref{sec:model}.
The implications of our model for the Majoron DM as its decay modes and production mechanisms, are shown in sections~\ref{sec:majdm} and \ref{sec:disc}. 
Finally, the conclusions are in section~\ref{sec:conc}.
\section{\label{sec:model} The Spontaneous Inverse Seesaw Model}
In this section we will describe the inverse seesaw (ISS) mechanism for neutrino mass generation~\cite{Mohapatra:1986bd}, with the addition of a spontaneously broken lepton number. 
In this context, $U(1)_{B-L}$ is taken as an \emph{ad-hoc} symmetry. 
Even though some efforts going in this way have been shown before (for instance at~\cite{Humbert:2015epa}), in this section we will describe the model and some consequences of the setup for the phenomenology.

\subsection{\label{sec:invseesaw} A Brief On Inverse Seesaw}

A remarkable feature of ISS is that it has different mass scales which configure the admixture of SM neutrinos with new fermion singlets.
There is a mass term, $m_D$, that connects SM neutrinos with the new singlets as a Dirac-like mass term. 
Another mass scale is represented by $M$ which gives the magnitude of the masses of the new singlets. 
Finally, a Majorana-like mass parameter $\mu$ connects the heavy and the light sectors after diagonalization of the mass matrix. 
In fact, the mass lagrangian for ISS can be written as~\cite{Abada:2014vea,Dias:2012xp,Garayoa:2006xs}:
\begin{eq}
  \label{eq:massnu}
  \mathcal{L} = - \frac{1}{2} n_L^T C \mathcal{M} n_L \, + \, h.c. \, ,
\end{eq}
where $n_L^T = (\nu_L, N_1^c, N_2)$ is composed by the SM neutrino $\nu_L$ and the new singlet fermions $N_{1,2}$.
For the case of 1 active neutrino and 2 singlets, the mass matrix $\mathcal{M}$ corresponds to a $3\times 3$ symmetric complex matrix given by
\begin{eq}
  \mathcal{M} &=& \left(\begin{array}{ccc}
                       0 & m_D e^{i \gamma_1} & 0 \\
                       m_D e^{i \gamma_1} & 0 & M e^{i \gamma_2} \\
                       0 & M e^{i \gamma_2} & \mu e^{i \gamma_3}
                      \end{array} \right) \, , \label{eq:massmat}
\end{eq}
phases $\gamma_i$ are added after allowing the mass terms to violate CP.
This matrix can be perturbatively diagonalized in a similar way to the Type-I seesaw when $\mu \ll m_D \ll M$, which in turn provides just one massive light neutrino.
Then the mass spectrum of the neutrino sector expanded at leading order in $m_D/M$ is
\begin{eq}
\label{eq:nueigenvals}
m_\nu &=& \left(\frac{m_D}{M}\right)^2 \mu ,\\
m_{\mathcal{N}_1} &=& M \,+\, \frac{m_D^2}{M} \,-\, \frac{\mu}{2}  \, , \\
m_{\mathcal{N}_2} &=& M \,+\, \frac{m_D^2}{M} \,+\, \frac{\mu}{2}  \, .
\end{eq}
For active neutrino masses of $m_\nu \sim$0.1~eV, heavy neutrinos with $M \sim$100~TeV and $m_D \sim$10~GeV, we require $\mu$ to be around 10~MeV~\cite{Dias:2012xp}, 
which matches our requirement for the hierarchy between the mass parameters. 
In that regime, the mixing matrix leading to the mass eigenstates reads
\begin{eq}
  U &=& \left(\begin{array}{ccc}
                       \displaystyle e^{-i\left(\gamma_1 - \gamma_2 + \frac{\gamma_3}{2}\right)} & \displaystyle 0 & \displaystyle - \frac{m_D}{M} e^{-i \frac{\gamma_3}{2}} \\[2ex]
                       \displaystyle \frac{i m_D}{\sqrt{2} M} e^{-i\left(\gamma_1 - \gamma_2 + \frac{\gamma_3}{2}\right)} & \displaystyle \frac{-i}{\sqrt{2}} e^{-i \left(\gamma_2 - \frac{\gamma_3}{2}\right)} & \displaystyle\frac{i}{\sqrt{2}} e^{-i \frac{\gamma_3}{2}} \\[2ex]
                      \displaystyle \frac{m_D}{\sqrt{2}M} e^{-i \left(\gamma_1 - \gamma_2 + \frac{\gamma_3}{2}\right)} & \displaystyle \frac{1}{\sqrt{2}} e^{-i \left(\gamma_2 - \frac{\gamma_3}{2}\right)} & \displaystyle \frac{1}{\sqrt{2}} e^{-i \frac{\gamma_3}{2}}
                      \end{array} \right) \, ,
\end{eq}
where the mass eigenstates are given by $ (\nu, \mathcal{N}_1, \mathcal{N}_2) =  \left( U\, n_L \right)^T$ and the mass matrix $\mathcal{M}$ is diagonalized by $m_{\nu}^{diag} = {\rm diag}\left( m_\nu, m_{\mathcal{N}_1}, m_{\mathcal{N}_2}  \right) = U \mathcal{M} U^T$.

\subsection{The Model \label{sec:spinvseesaw}}
The mass parameters in the inverse seesaw can be generated by means of spontaneous symmetry breaking (SSB) of a global $U_{l}(1)$ symmetry associated to the lepton number (see, for instance \cite{Humbert:2015epa,Humbert:2015yva}).
Our approach uses the following lagrangian:
\begin{eq}
  \label{eq:slangran}
  \mathcal{L} = - y_{L} e^{i \phi_L} \bar{L} H N_1^c - y_S e^{i \phi_S} S^\dagger \overline{N_2} N_1^c - \frac{y_X e^{i \phi_X}}{2}  X^\dagger \overline{N_2^c} N_2 + h.c. \, \label{eq:lagISS},
\end{eq}
The Higgs doublet is defined by $H^T = \left( \chi^+, \left(v_h + \sigma_h + i \chi_h\right)/\sqrt{2} \right)$ where $\sigma_h (\chi_h)$ is the (pseudo)scalar component of the Higgs doublet whose vev is $v_h \simeq 246$~GeV, while $\chi^{+}$ is the longitudinal component of the $W^+$.
We have included two complex scalars, $S$ and $X$, which are charged with lepton number although they are neutral under the whole SM gauge group. After SSB, these fields acquire complex vevs named $v_S e^{i \theta}$ and $v_X e^{i \tau}$ (see Section~\ref{sec:scalpot}). 
Notice that CP violating couplings and vevs have been included for the sake of generality. 
We will also include phases for the $N_j$'s as a way to reabsorb as many phases in the couplings as possible, thus: $N_j \rightarrow e^{i \psi_j} N_j$. After these inclusions, the mass 
parameters can be defined by:
\begin{eq}
  m_D e^{i \gamma_1} &=& \frac{y_L v_h}{\sqrt{2}} e^{i (\phi_L - \psi_1)} \, , \\
  M e^{i \gamma_2} &=& \frac{y_{S} v_S}{\sqrt{2}} e^{i (\phi_S - \psi_1 - \psi_2 - \theta)} \, , \\
  \mu e^{i \gamma_3} &=& \frac{y_{X} v_X}{\sqrt{2}} e^{i (\phi_{X} + 2 \psi_{2} - \tau)}\, ,
\end{eq}
By demanding the moduli $M$ and $\mu$ to be 100 TeV and 10 MeV respectively, and with yukawa couplings that cannot exceed the perturbative limit, we may set bounds for the moduli of the vevs of $S$ and $X$, namely, $v_S$ and $v_X$:
\begin{eq}
v_S > \frac{M}{\sqrt{2 \pi}} \, , \\
v_X > \frac{\mu}{\sqrt{2 \pi}} \, ,
\end{eq}
which corresponds to $v_S > 50$~TeV and $v_X > 5$~MeV. On the contrary, the value of $m_D$ is completely fixed by $y_L$ since $v_h$ has a well-defined value.

On the other hand, the $U(1)_l$ charges have been assigned by requiring Eq.~\ref{eq:slangran} to be lepton number invariant. The resulting charge assignment is shown in Table~\ref{tab:charges}. 
Note that not all the charges can be fixed by Eq.~\ref{eq:slangran}, which leaves the assignments of $N_2$, $S$ and $X$ as functions of the lepton number of $N_2$, which we call $x$. 
The value that this charge will take and the fate of the CP violating phases, will be depicted in the following sections.

\begin{table}[tb]
\centering
\begin{tabular}{|c|ccccc|}
\hline
            &  $L$  &  $N_1$  &  $N_2$  &  $S$  &  $X$  \\
\hline\hline
 $SU(2)_{L}$  & $2$   & $1$ & $1$  & $1$   & $1$ \\
\hline
$U(1)_{Y}$  & $1/2$ & $0$ & $0$  & $0$   & $0$ \\
\hline
$U(1)_{l}$  & $1$   & $-1$ & $x$ & $1-x$ & $2x$ \\
\hline
\end{tabular}
\caption{\label{tab:charges} Charge assignment of the model. }
\end{table}

\subsection{\label{sec:scalpot} The Scalar Potential} 

The scalar potential for the new singlets $S$ and $X$ is given by
\begin{eq}
  \label{eq:potsx}
  V_{SX} = -\mu_{S}^2 \left|S\right|^2 + \frac{\lambda_S}{4} \left|S\right|^4 - \mu_{X}^2 \left|X\right|^2 + \frac{\lambda_X}{4} \left|X\right|^4 + \lambda_5 \left|S\right|^2 \left|X\right|^2 + V_{\rm I} \, ,
\end{eq}
where $\mu_i^2$ are positive mass terms, $\lambda_i$ are adimensional couplings allowed by perturbative limit. However, the chage assignment we showed at equation~\ref{tab:charges} allows an additional term which we encode within $V_I$
\begin{eq}
  \label{eq:potvi}
  V_{\rm I} = \lambda_{J} e^{i\delta} X {S^{\dagger}}^3 + {\rm h.c.} \, \label{eq:vcp},
\end{eq}
Notice that after rephasing the fields $S$ and $X$, this term is not necessarily CP invariant. 
In fact, even after promoting the coupling $\lambda_J$ to explicitly have a complex phase, that doesn't ensure that the Lagrangian is CP invariant, since the rephasing of the fields will pop out in the Yukawa interactions involving $N_i$'s. 
Nevertheless, if we demand this term to be lepton number invariant we will 
fix the lepton charge $x$\footnote{Different models with a similar underlying idea 
can be found at~\cite{Berezinsky:1993fm,Chulia:2016giq,Gu:2010ys,Aranda:2009yb}}. 
For this particular case, the charge assignments are taken as $L_{N_1} = 1$, $L_{N_2} = x = 3/5$, 
$L_X = 2x = 6/5$, and $L_S = 1-x = 2/5$.

The remaining terms of the scalar potential include the Higgs piece
\begin{eq}
  V_{\rm HSX} = -\mu_{H}^2 H^{\dagger}H + \frac{\lambda_{H}}{4}(H^{\dagger}H)^2 + \lambda_{HS} \left|S\right|^2 H^{\dagger}H + \lambda_{HX} \left|X\right|^2 H^{\dagger}H \, , \label{eq:higgsmix}
\end{eq}
where $\mu_H^2$ and $\lambda_H$ are the Higgs mass parameter and its quartic self-interaction, 
while $\lambda_{HS}$ and $\lambda_{HX}$ are the couplings between $H$ and the new scalars. Therefore,
the full scalar potential is the sum of Eqs.~\ref{eq:potsx} and \ref{eq:higgsmix},
\begin{eq}
\label{eq:vscal}
V_{\rm scalar} = V_{SX} + V_{HSX} \, .
\end{eq}

After SSB, the fields $S$ and $X$ can be expanded around a non CP-invariant vacuum
\begin{eq}
  S &=& \frac{e^{i\theta}}{\sqrt{2}} \left(v_S + \sigma_S\, + i\chi_S \right) \label{eq:seps} \, \\
  X &=& \frac{e^{i\tau}}{\sqrt{2}} \left(v_X + \sigma_X + i\chi_X \right)  \label{eq:sepx} \, ,
\end{eq}
where $\theta$ and $\tau$ are phases already introduced in Section~\ref{sec:spinvseesaw}. The 
physical fields are extracted after the minimization of the $V_{\rm scalar}$ and plugging 
back into the Lagrangian the solutions of the tadpole equations
\begin{eq}
\left. \frac{\partial V_{\rm scalar}}{\partial s_0^i} \right|_{s_0^{j\neq i} =0} = 0 \, \label{eq:tadpoles},
\end{eq}
where ${s^0}^T = \left( \sigma_S,\sigma_X,\sigma_h,\chi_S,\chi_X \right)$. Afterwards, 
we still need to write down and diagonalize the mass matrices in order to get the physical 
fields, something we will examine in the next section. However, it is remarkable that the 
tadpole equations for the fields $\chi^{+}$ and $\chi_h$ are trivially satisfied and do not 
add other relevant information than they are taking part as the longitudinal components of 
the gauge bosons $W$ and $Z$. Hence, the equations ~\ref{eq:tadpoles} give rise to the following 
relations among the parameters
\begin{eq}
\label{eq:relationtadpole}
\tau    &=& 3\theta - \delta - \pi \, ,\\
\mu_S^2 &=& \frac{v_S^2}{4} \left( 2 \epsilon_h^2 \lambda_{HS} + \lambda_S  - 6 \lambda_{J} \omega + 2 \lambda_5 \omega^2 \right) \, , \\
\mu_X^2 &=& \frac{v_S^2}{4} \left( 2 \epsilon_h^2 \lambda_{HX} - 2 \lambda_{J}\, \omega^{-1} + 2 \lambda_5  + \lambda_X \omega^2\right) \, , \\
\mu_H^2 &=& \frac{v_S^2}{4}  \left( \epsilon_h^2 \lambda_H + 2 \lambda_{HS} + 2 \lambda_{HX} \omega^2 \right) \, ,
\end{eq}
where $\omega = v_X/v_S$ and $\epsilon_h = v_h/v_S$. Since the value of $v_S$ has a lower limit
of around $50$ TeV, the ratio $\epsilon$ turns out to be at least $5 \times 10^{-3}$. Even though
there is a similar lower limit for $v_X$, there is no particular value for $\omega$ that we could 
take a priori. 

Regarding the first of the equations \ref{eq:relationtadpole}, we may draw the following relations 
among CP-phases present in the scalar and neutrino sectors,
\begin{eq}
  \gamma_1 &=& \phi_L - \psi_1 \, , \label{eq:gamma1}\\
  \gamma_2 &=& \phi_S - \psi_1 - \psi_2 -\theta \, ,\label{eq:gamma2}\\
  \gamma_3 &=& \phi_X - 2 \psi_2 - \tau \, ,\label{eq:gamma3}
\end{eq}
and Equation~\ref{eq:relationtadpole}. In our particular setup of ($1\nu$, $2N$) neutrinos, 
$\gamma_i$ phases can be rotated away by the field phases $\psi_i$, $\theta$, and $\tau$, obtaining 
thus a neutrino mass matrix $\mathcal{M}$ which is real and symmetric. This means even though it
was apparent that CP could be violated in this model, there is no source of CP violation across the
Lagrangian. However in an extended scenario with more generations of active and singlet neutrinos, 
we will get CP violation in the neutrino sector due to the imposibility of absorbing all phases. 
Similar situation occurs in the scalar sector when extra copies of $S$ are added~\cite{Geng:1988gr}.
Nonetheless, we will stick throughout this work to a model having only one active neutrino and two 
fermion singlets.

\subsection{\label{sec:spectrum} Mass Spectrum}
The mass matrix for the scalar and pseudoscalar fields is obtained from
\begin{eq}
\frac{\partial^2 V_{\rm scalar}}{\partial s^0_i \partial s^0_j} \equiv \left\{M^2\right\}_{ij} \, \label{eq:scalmss}.
\end{eq}
Unfortunately, the phases we included spureously mix up these two sectors, but since
the Lagrangian we have chosen conserves CP, we can safely rotate away this admixture.
On top of that, the admixture between $\chi^{h}$ and the rest of the scalars is identically 
zero, so this field is not included on \ref{eq:scalmss}. After proceeding with the derivatives
of the scalar potential, one can transform the resulting mass matrix by means of the
following rotation matrix,
\begin{eq}
R_{\rm cp} = \left( \begin{array}{ccccc}
           c_\theta & 0 & 0 & s_\theta & 0 \\
           0 & c_{\delta - 3\theta} & 0 & 0 & s_{\delta - 3\theta} \\
           0 & 0 & 1 & 0 & 0 \\
           -s_\theta & 0 & 0 & c_\theta & 0 \\
           0 & -s_{\delta - 3\theta} & 0 & 0 & c_{\delta - 3\theta}
               \end{array} \right)  \, ,
\end{eq}
which acting upon $M^2$ produces the following block diagonal form,
\begin{eq}
  M_{s}^2 = R_{\rm cp} M^2 R_{\rm cp}^T =
    \left( \begin{array}{cc}
            M_{\rm scal}^2 & 0 \\
            0 & M_{\rm pscal}^2
            \end{array} \right)   \, ,
\end{eq}
where
\begin{eq}
M_{\rm scal}^2 &=& \frac{v_S^2}{2}
\left( \begin{array}{ccc}
\lambda_S - 3\lambda_{J}\omega  & 2\lambda_5\omega - 3\lambda_{J} & 2 \epsilon_h \lambda_{HS}  \\
2\lambda_5\omega - 3\lambda_{J} & \displaystyle{\frac{(\lambda_X\omega^3 + \lambda_{J})}{\omega}}  & 2 \epsilon_h \lambda_{HX} \omega \\
2 \epsilon_h \lambda_{HS} & 2 \epsilon_h \lambda_{HX} \omega & \epsilon_h^2 \lambda_H \\
\end{array} \right)  \label{eq:matscal} \\
M_{\rm pscal}^2 &=& \frac{v_S^2}{2} \, \lambda_{J} \,
\left( \begin{array}{cc}
               9 \omega & -3  \\
               -3 & \omega^{-1}
              \end{array} \right) \, .
\end{eq}

Which means that the scalar and pseudoscalar blocks are now disentangled and
both matrices can be diagonalized separately. Thus, the model has 5 mass 
eigenstates labeled as $\zeta_i, i=1\dots5$. The first two (i.e. $\zeta_1$ and 
$\zeta_2$) correspond to those from $M^2_{\rm pscal}$, the rest comes from  
$M^2_{\rm scal}$ where $\zeta_5$ is reserved to the SM-like higgs with a mass 
of $m_{\zeta_5} = m_{h} = 125$~GeV.

The two eigenstates of the matrix $M^2_{\rm pscal}$ are obtained after applying
the diagonalization $R_{\rm pscal} M_{\rm pscal}^2 R_{\rm pscal}^T = {\rm diag}(m_{\zeta_1}^2, m_{\zeta_2}^2)$
where
\begin{eq}
  R_{\rm pscal} = \frac{1}{\sqrt{1 + 9 \omega^2}}\left( \begin{array}{cc}
    1 & 3 \omega \\
    -3 \omega & 1
  \end{array} \right) \, .
\end{eq}
The $\zeta_1$ state corresponds to the Nambu-Goldstone boson of the $U(1)_l$ 
breaking and it is the commonly known {\it massless} Majoron. Whereas the 
second eigenstate has a mass of
\begin{eq}
m^2_{\zeta_2} = M_J^2 = \frac{v_S^2}{2 \omega} \lambda_{J} (1 + 9 \omega^2) \label{eq:majmass} \, ,
\end{eq}
and it can be thought as a {\it massive} Majoron. This state will be considered 
as the DM candidate of the model and we label it as $\zeta_2 = J_{\rm DM}$. Even though
the \emph{massless} Majoron is commonly called as the Majoron-DM, in this setup
we will fully disregard about its potential role as a DM candidate. 

There is only one parameter at the scalar potential controlling $M_J$, namely, $\lambda_J$. 
This mass is not arbitrary in our setup and it will be strongly fixed by the DM half-life. As we
shall see, in order to have a cosmologically stable DM candidate, we need to have $M_J\sim 10$~keV
as upper bound. This immediately implies a bound for $\lambda_{J}$ which is $\simeq \frac{M_J^2}{v_S^2} < 10^{-22}$.
This is mainly given by the enormous lower limit of $v_S$ and for the smallness of $M_J$.
Even though this sounds a little bit extreme, there are two good points regarding this
suppression. On the one hand, this limit implies kinematical suppression for many 
possible decay channels. On the other hand, as it is shown in the appendix \ref{sec:2loop},
the running of $\lambda_J$ is proportional to itself, so that this coupling can be 
safely taken small and quantum corrections do not spoil this smallness.

On different grounds, the scalar mass matrix $M_{\rm scal}^2$provides 3 massive states,
which can be found by means of a perturbative diagonalization by using $\lambda_{J}$ 
and $\epsilon_h$ as perturbative parameters. In that setup, the SM-like higgs has 
a mass of
\begin{eq}
\label{eq:hmass}
m_{h}^2 \simeq \frac{v_h^2}{2} \, \left\{ \frac{\lambda_H}{2} + 2 \left(\frac{ \lambda_{HX}^2 \lambda_S +
  \lambda_{HS}^2 \lambda_X -4 \lambda_5 \lambda_{HS} \lambda_{HX}}{4 \lambda_5^2 - \lambda_S \lambda_X} \right) \right\} \, .
\end{eq}
The constraint regarding the higgs mass, $m_h = 125.18 \pm 0.16$~GeV~\cite{PhysRevD.98.030001}, 
provides restrictions on the values of the couplings in Eq.~\ref{eq:hmass}. At the same time, 
in order to ensure that the Higgs we are considering is mostly SM-doublet-like,
the scalar mixing matrix is constrained to have the element $U_{h5} > 0.86$~\cite{Cheung:2015dta}. 
This value allows only a small portion of $\sigma_{S,X}$ to be present in $\zeta_5$. 
In the perturbative diagonalization, $\zeta_5$ is written as combination of $\sigma_{h,S,X}$
\begin{eq}
  \zeta_5 \simeq \sigma_h + 4 \epsilon_h \left( \frac{2 \lambda_{HX} |\lambda_5| + \lambda_{HS} \lambda_X}{4 \lambda_{5}^2 - \lambda_{S}\lambda_{X}}  \sigma_S
+ \frac{2 \lambda_{HS} |\lambda_5| + \lambda_{HX}\lambda_S}{4 \lambda_{5}^2 - \lambda_{S}\lambda_{X}}  \sigma_X \right) \, ,
\end{eq}
where the portion related to $\sigma_{S,X}$ is, in general, suppressed by terms 
$\mathcal{O}(\epsilon_h)$ and thus the $\zeta_5$ is mostly SM-like Higgs. Nevertheless,
this suppression is not present for $m_h^2$, as it can be read from~\ref{eq:hmass},
and thus large contributions from $\lambda_{HS,HX}$ might arise. In order to stick
the Higgs mass to have a form closer to $m_{h}^2 \simeq \frac{\lambda_H v_h^2}{4}$,
we will use the following limit $\lambda_{HX}$, $\lambda_{HS} \ll 1$.

In that perturbative scheme, the remaining two massive states ($\zeta_{3,4}$) have masses:
\begin{eq}
M_{\zeta_3}^2 &\simeq& \frac{v_S^2}{2} \left( \frac{-A\,+\,A\psi\, +\, 2\lambda_X\omega \psi}{2\psi} \right) \, , \\
M_{\zeta_4}^2 &\simeq& \frac{v_S^2}{2} \left( \frac{ A\,+\,A\psi\, +\, 2\lambda_X\omega \psi}{2\psi} \right) \, ,
\end{eq}
where the parameters $A$ and $\psi$ have been defined by means of
\begin{eq}
\lambda_S &=& A\, +\, \lambda_X \omega^2 \, ,\\
\lambda_5 &=& -A\, \left( \frac{\sqrt{1 - \psi^2}}{ 4\, \omega\, \psi} \right) \, .
\end{eq}
The $A$ parameter can be seen as an alignment between $\lambda_S$ and $\lambda_X$ and 
therefore it has a range value of a typical adimensional coupling. Since nothing is imposed 
upon $A$ and $\psi$, we expect $M_{\zeta_3, \zeta_4}$ to be in the order of $\mathcal{O}\left(v_S\right)$.
The $\psi$ term is $\cos{\left(\phi/2 \right)}$ where $\phi$ is the mixing angle of the 
heavy sector in the matrix $M_{\rm scal}^2$sector $\sigma_S$ and $\sigma_X$. Since it 
is a trigonometric function, $\lambda_5$ can take positive and negative values.

As a summary for this section, the mass spectrum of the scalar and pseudoscalar sector 
has 3 well defined mass scales. First, we have the light states around $\sim \mathcal{O}({\rm keV})$,
corresponding to the scale for massive Majorons. After that we have the scale for Higgs mass
which is set at $\sim 125$~GeV. Finally we have the scale for the heavy scalars given by $v_S$, which
goes above the $10^2$~TeV.

\section{\label{sec:majdm}Majoron Dark Matter}

As it was indicated in the previous section, our dark matter candidate corresponds 
to the {\it massive} Majoron $J_{\rm DM}$. In this context, and since it
is massive, this particle will be a decaying DM candidate~\cite{Aranda:2009yb,Akhmedov:1992hi,Lattanzi:2014mia, Adhikari:2016bei} 
where its decay channels are mainly to neutrinos and {\it massless} Majorons.
In the previous section we advanced that this is the limit we will take through
this work. A different channel that can be activated if we take 
the Majoron mass slightly above $m_h$ is $J \longrightarrow \zeta_5 \zeta_1$.
In turn, that channel will be shut if we take its mass slightly below $m_h$, and in our
case, the DM cosmological stability asks to have a Majoron with a mass many orders
of magnitude away below that limit, in the scale of a few keV's. 
In this section, we focus on the decay modes to $2\nu$, $2\zeta_1$ and $3\zeta_1$, and also 
we comment on the DM production in the Early Universe.\\

\subsection{\label{sec:decay}Dark Matter Decay}

In the case of decaying DM, the main phenomenological constraint comes from 
the cosmological stability. Therefore, we will assume in our case that $J_{\rm DM}$ 
has a lifetime $\tau_{\rm DM} > 10^{19}$~s ($\Gamma_{\rm DM} < 10^{-44}$~GeV)~\cite{Lattanzi:2013uza}.
This limit has been taken since the new physics in the scalar sector does not 
involve charged particles linked to either $S$ or $X$, and thus decays of DM into 
photons are forbidden. However, there are constructions that allow this 
mode\cite{Lattanzi:2007ux}. Otherwise, a much more stringent bound should be used~\cite{Cirelli:2012ut}.
In this model, we have two classes of decay modes: fermionic ($2\nu$) and scalar 
($2\zeta_1$ and $3\zeta_1$). The first one is the typical Majoron signature~\cite{Aranda:2009yb,Akhmedov:1992hi,Lattanzi:2014mia,Schechter:1981cv}
that is already present in other Majoron DM models given by its lepton-number-breaking 
nature. The second class corresponds to scalar modes coming from the potential (Eq.~\ref{eq:vscal}).

\subsubsection{Decay into neutrinos}
The decay rate $J_{\rm DM} \rightarrow \nu \nu $ in the limit $m_\nu \ll M_J$ is
\begin{eq}
\label{eq:nudecay}
\Gamma_\nu \simeq \frac{M_J}{32\pi} \left( \left|\left| O_L \right|\right|^2\,+\, \left|\left| O_R \right|\right|^2\right)  \, .
\end{eq}
The couplings $O_L$ and $O_R$ are the couplings of Majoron to neutrinos,
and they are given in the appendix~\ref{app:fercoup}. The decay rate coming
from this process can be rewritten as
\begin{eq}
\label{eq:J2nu}
\Gamma_\nu = \frac{M_J}{32\pi} f\left( m_\nu,m_D,M,v_S\right) \, ,
\end{eq}
where the function $f$ is described in appendix~\ref{app:fercoup} and it 
contains the dependence on the parameters of the couplings between neutrinos 
and the majoron DM. With these elements, the decay rate can be expanded in 
powers of $\displaystyle \frac{\mu}{M} = \alpha \sim 10^{-7}$, which up to 
order $\alpha$ is
\begin{eq}
\label{eq:gammanu}
\Gamma_\nu = \Gamma_{0\nu}(\omega) \, \left\{(2-3\omega^2)\left(2-3\omega^2(1+2 \alpha)\right)  + \mathcal{O}(\alpha^2) \right\}\, ,
\end{eq}
where the overall factor is:
\begin{eq}
\Gamma_{0\nu}(\omega) = \frac{M_J {m_{\nu}}^2}{256 \pi v_S^2} \, \frac{1}{\omega^2 (1 + 9 \omega^2)} \, .
\end{eq}
Our strategy is to look for a parameter space that allows a suppression for this decay rate,
which will be neat if we vanish the equation~\ref{eq:gammanu} up to first order in $\alpha$.
Indeed, this is rather easily given by taking $\omega = \omega_0 = \sqrt{2/3}$. The error 
carried by this choice produces a decay rate of $\Gamma_\nu = \Gamma_{0\nu}(\omega_0)\, 4 \alpha^2$.
This value for $\omega$ suggests that $v_X$ and $v_S$ should be around the same
order of magnitude, and further powers of $\alpha$ will act as perturbations around 
that value for $\omega_0$.

The overall factor $\Gamma_{0\nu}$ can be evaluated at $\omega_0$, in order to explore
the size of the decay rate
\begin{eq}
\Gamma_{0\nu}(\omega_0) \simeq 10^{-42} \, {\rm GeV} \left(\frac{m_{\nu}}{0.1\,{\rm eV}} \right)^{2} \left(\frac{M_J}{1 \, {\rm keV}}\right) \left(\frac{v_S}{10^6\,{\rm GeV}}\right)^{-2} \, .
\end{eq}
Recalling that the cosmological stability requires a lifetime larger than $10^{-44}$ GeV,
and given that the decay rate is given by $\Gamma_\nu = \Gamma_{0\nu}(\omega_0)\, 4 \alpha^2$,
with $\alpha^2 \sim 10^{-14}$, it is ensured that $\Gamma_{\nu} \sim 10^{-58}$ GeV for a
keV Majoron. Since the suppression in this channel is given mainly by the smallness of neutrino
masses and the values of the inverse seesaw parameters when evaluating $\omega \sim \omega_0$, the
decay to neutrinos will not put constraints on $M_J$.

Nonetheless, in this work we have taken the exact solution for $\Gamma_\nu = 0$ at all 
orders in $\alpha$ in order to see how large are the variations of $\omega$ around
$\omega_0=\sqrt{2/3}$. The full solution for $\omega$ is
\begin{eq}
\label{eq:omega23}
\omega = \frac{\sqrt{4 - 4 \alpha^2 + \alpha^3 - \alpha^4}}{\sqrt{3(2 + 2 \alpha + \alpha^4)}}.
\end{eq}

This value moves around $\omega_0$ with fluctuations in the order of $\alpha\sim\mu/M$. It 
is quite remarkable that the $\mu/M$ ratio is at the same time responsible for suppression
for neutrino masses in the inverse seesaw mechanism (see ~\ref{eq:nueigenvals}), and a pivotal
element for DM stability.

\subsubsection{Decay Into Scalars}
As it was advanced, the scalar decay modes for the massive Majoron can only be $J_{\rm DM} \rightarrow 2 \zeta_1$
and $J_{\rm DM} \rightarrow 3 \zeta_1$ when $M_J$ ranges between $\mathcal{O}(1)$ keV up to  $\mathcal{O}(1)$ 
GeV. These are the channels we will focus on in this subsection. It is worth recalling that we will consider
also that $M_{\zeta_3,\zeta_4}$ have masses far larger than $m_h$, and thus, no other channels will be open.

The couplings required to calculate the decay rate come from the scalar potential,
and they are obtained after writing it in terms of the mass eigenstates. Subsequently, 
depending on the number of legs of the interactions, derivatives are taken in the
following way
\begin{eq}
\lambda_{ijk} = \frac{\partial^3 V_{\rm scalar}}{\partial \zeta_i \partial \zeta_j \partial \zeta_k} \quad {\rm and} \quad  \lambda_{ijkl} = \frac{\partial^4 V_{\rm scalar}}{\partial \zeta_i \partial \zeta_j \partial \zeta_k \partial \zeta_l} \, .
\end{eq}

The 2-body decay mode $J\longrightarrow2\zeta_1$ requests to know the coupling $\lambda_{211}$.
After computing the derivatives of the scalar potential, we obtain $\lambda_{211} = 0$. Thus,
even though this mode should be dominant, it is not present in our model regardless any
consideration made upon the values of the couplings. This is mainly given as a consequence
of CP-invariance, which is conserved in the Lagrangian as it was shown in the section
\ref{sec:scalpot}. Henceforth, we will compute the contributions coming only from $J\longrightarrow3\zeta_1$.

\begin{figure}[t]
\includegraphics[width=0.7\textwidth]{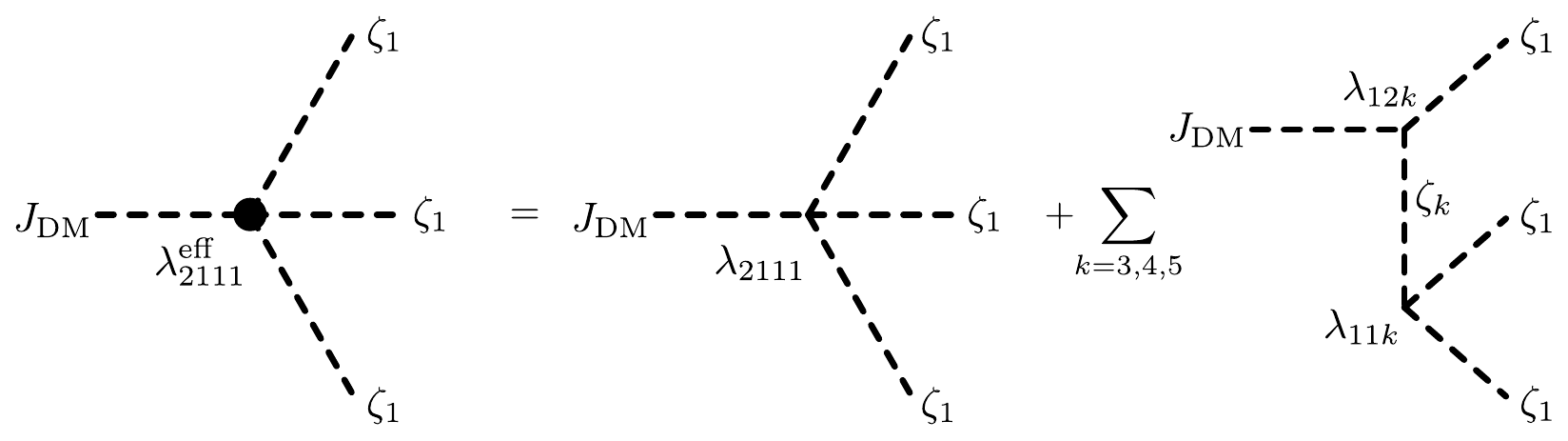}
\caption{\label{fig:lambeff} Diagrams involve in the calculation of $\lambda_{2111}$ for the process $J_{\rm DM} \rightarrow 3 \zeta_1$}
\end{figure}

The decay rate for the 3-body decay considers contributions from many diagrams. The process 
can be calculated as
\begin{eq}
\Gamma_{3\zeta} &=& \frac{1}{(64\pi)^3} M_J \left|\left| \lambda^{\rm eff}_{2111} \right|\right|^2 \label{eq:3zeta}
\end{eq}
where $\lambda^{\rm eff}_{2111}$ includes the graphs shown in Fig.~\ref{fig:lambeff}.

Then, we have two sources of contributions. One, is a cuartic interaction between $\zeta_2$ and three
$\zeta_1$ (which is dubbed as $\lambda_{2111}$). The other contributions are mediated by the scalars 
$\zeta_3$ and $\zeta_4$ and $\zeta_5$. All of them are considered heavy particles since we are
sticked to the limit $M_J\sim \mathcal{O}(1)$~keV (see section \ref{sec:spectrum}). Within this limit, 
the effect of the three scalar mediators can be safely integrated out\footnote{This comes from the 
fact that in the Majoron's restframe, it cannot decay into particles heavier than itself, 
like $\zeta_3$, $\zeta_4$ and $\zeta_5$. Hence, only off-shell production of the heavy particles 
and its subsequent decay are allowed. The larger is difference in mass, the better this 
approximation will work.}, giving rise to
\begin{eq}
\label{eq:scaleff}
\lambda^{\rm eff}_{2111} &=& \lambda_{2111}\,-\,\frac{\lambda_{213}\lambda_{113}}{m_3^2}\,-\,\frac{\lambda_{214}\lambda_{114}}{m_4^2}\,-\,\frac{\lambda_{215}\lambda_{115}}{m_5^2}  \, ,
\end{eq}
where the full expressions for $\lambda_{2ij}$ and $\lambda_{2111}$ are shown in the appendix \ref{app:scacoup}.
Recall that these couplings have different mass dimensions each.

Our strategy will be similar to that we used to look for a parameter space that minimizes the 
decay mode $J\longrightarrow 2\nu$, i.e. the aim is to probe it via imposing $\lambda^{\rm eff}_{2111} \simeq 0$.
In the practice, the combination $\lambda_{2111}\,-\,\frac{\lambda_{213}\lambda_{113}}{m_3^2}\,-\,\frac{\lambda_{214}\lambda_{114}}{m_4^2}$ 
is mainly composed by unsuppressed couplings plus corrections proportional to $\left(M_J/v_S \right)^2$. 
On the other hand, the coupling related to Higgs exchange (i.e. $\frac{\lambda_{215}\lambda_{115}}{m_5^2}$) 
is proportional to an overall factor of $\left(M_J/v_S \right)^4$ (see appendix \ref{app:scacoup}), since
the only source of this couplings is via projecting the scalar parts of $S$ and $X$ onto the Higgs via
equation~\ref{eq:vcp}. 

The factor $M_J/v_S$ is $\mathcal{O}\sim 10^{-11}$ for a keV Majoron. Different
from the case of the decay to neutrinos, this ratio will be in charge of suppressing 
at least a part of the contribution to $J \longrightarrow 3\zeta_1$. We will aim to get 
$\lambda_{2111}\,-\,\frac{\lambda_{213}\lambda_{113}}{m_3^2}\,-\,\frac{\lambda_{214}\lambda_{114}}{m_4^2} \sim \left(M_J/v_S \right)^2$.
Other ways to suppress this channel are of course allowed, but the algebraic form of the couplings involved
ends up being difficult to deal with. Just in order to fix some ideas, let's evaluate 
the decay  $J\longrightarrow 3\zeta_1$ when only the Higgs contribution is considered:

\begin{eq}
\Gamma_{3\zeta_1} &=&  \frac{1}{(64\pi)^3}\, M_J \left|\left| \frac{\lambda_{215}\lambda_{115}}{m_5^2} \right|\right|^2  \\
         &=& \frac{1}{(64\pi)^3}\, M_J\, \left( \frac{M_J}{v_S} \right)^8\cdot \left( \frac{\psi^2 \omega^3}{(1 + 9 \omega^2)^3}\,\mathcal{F}\left(A,\psi,\omega,\lambda_h,\lambda_{HS},\lambda_{HX}\right) \right)^2 \, ,\nonumber
\end{eq}
where the function $\mathcal{F}\left(A,\psi,\omega,\lambda_h,\lambda_{HS},\lambda_{HX}\right)$ 
goes to zero when both couplings $\lambda_{HS}$ and $\lambda_{HX}$ go to zero. For a wide 
range of the values of the couplings, we obtain that $\mathcal{F} \sim \mathcal{O}(1)$.
This implies that the decay rate is suppressed mainly by $\left( M_J/v_S \right)^8$.
Hence, an orders of magnitude evaluation of the decay rate, for a keV Majoron, is given by:
\begin{eq}
\Gamma_{3\zeta_1}^{\zeta_5} \sim 10^{-12} \, \left(\frac{M_J}{1 \, {\rm keV}}\right) \left( \frac{M_J}{v_S} \right)^8\, {\rm GeV}\,\, \sim \,\, 10^{-108}\, {\rm GeV} \, . \label{eq:higgsmed}
\end{eq}

Thus the channel to Higgs is extremely suppressed. Unfortunately, this suppression is 
way less extreme in the case of $\lambda_{2111}\,-\,\frac{\lambda_{213}\lambda_{113}}{m_3^2}\,-\,\frac{\lambda_{214}\lambda_{114}}{m_4^2}$.
After grouping the contributions of this combination in powers of $M_J/v_S$ we get 
that the decay rate proportional to $(M_J/v_S)^4$, which ends up being larger than
the contribution of Higgs. Thus an orders of magnitude evaluation of this piece of the channel
gives rise to
\begin{eq}
\Gamma_{3\zeta_1}^{\neq \zeta_5} \sim 10^{-12} \, \left(\frac{M_J}{1 \, {\rm keV}}\right) \left( \frac{M_J}{v_S} \right)^4\, {\rm GeV}\,\, \sim \,\, 10^{-52}\, {\rm GeV} \, .
\end{eq}

If we use a MeV massive Majoron, we will have $\Gamma_{3\zeta_1}^{\neq \zeta_5} \sim 10^{-41}$ GeV,
which means we are off by 3 orders of magnitude. However, by using a $100$~keV Majoron 
we will get $\sim 10^{-46}$ GeV, which is allowed by 2 orders of magnitude and it can be 
considered as an upper bound for Majoron mass. As it was advanced, this limit can be 
relaxed by studying the full parameter space for which $\left|\left| \lambda_{2111}^{eff}\right|\right| = 0$. 
Unfortunately, the larger the Majoron 
mass becomes, the larger the number of decay channels gets. This issue addresses that
it could be more difficult to find combinations of parameters for a cosmologically stable 
Majoron if its mass is increased.

Nonetheless, keeping the Majoron below the ~keV limit implies 
an interesting property apart from shutting down some decay channels. In fact, that points towards
a slightly broken global symmetry that charges the fields $S$ and $X$, whose breaking interaction
is given by $\lambda_J$. Under this description, that symmetry fits well
with a new $U(1)$, and it cannot be identified with lepton number since this number is 
conserved with our charge assignment (as it shown in the appendix \ref{app:charge}).
The latter point is reinforced by looking at the 2 loop RGE for the coupling $\lambda_J$
(see appendix \ref{sec:2loop}), which goes to zero if that parameter is set initially 
to zero. These two points are enough to justify the selection for DM mass at this low
scale, which will be the setup we will stick to in this work.

On the other hand, notice that even though we need $\lambda_{HS}$ and $\lambda_{HX}$ to be small
in order to ensure a Higgs mass mainly given by the SM Higgs doublet,
we have not applied this limit through this section. We only restricted ourselves
to examine the suppressions appearing with powers of $M_J/v_S$,
which by its own ensures a cosmological half-life for our DM candidate. However, 
it suffices to put $\lambda_{HS},\lambda_{HX} \sim 10^{-3}$
in order to ensure small mass mixing between Higgs and the heavy scalars. For the 
sake of numerical examination of a parameter space made of 8 quantities ($\lambda_X$, 
$\lambda_5$, $A$ and $\psi$ in the scalar sector and $M$, $m_D$, $\mu$
and $\omega$ in the fermion sector), in the following sections we will 
consider both couplings to be zero, which is a harmless selection neither for DM half 
life nor the DM production mechanism. A more involved examination will require to 
rise that restriction, indeed. The details of this calculation 
will be given at section \ref{sec:disc}.

\subsection{Dark Matter Production}

In this section, we aim to describe a tentative framework for the DM production in 
the Early Universe. Our DM candidate $J_{\rm DM}$ shares similar properties with a 
Feebly Interacting Massive Particle (FIMP)~\cite{Hall:2009bx} by means of suppressed 
coupling with the SM-like higgs and active neutrinos. However, the couplings of $J_{\rm DM}$ 
to the heavy scalars $\zeta_3$ and $\zeta_4$ may not be necessarily suppressed, and 
in turn, the couplings of these particles with the SM-like Higgs might take a wide 
range of values. This makes the heavy scalars be able to interact with the rest 
of the thermal bath, and subsequently decay to $J_{\rm DM}$.

Indeed, the same logic could be applied to the process involving heavy neutrinos.
Under these conditions, and assumming a keV DM candidate, the production mechanism 
cannot be addressed with the typical freeze-out for $J_{\rm DM}$, which is used 
in WIMP-DM models to reproduce the relic abundance~\cite{Gondolo:1990dk}.

On top of having $J_{\rm DM}$ as a FIMP-DM, the Lightest Observable Sector Particle 
(LOSP) could be either the lighest of the heavy neutrinos or the lightest between 
$\zeta_{3,4}$ because all of them have $U(1)_l$ charges.
Due to the interplay among all particles of the model, the relic abundance calculation 
has many edges which at first sight are not evident to spot (\cite{Heeck:2017xbu} presents 
a very clear explanation on the relevant observables). Despite this difficulty, we will sketch some 
relevant processes involved in the DM production.

Some of the prototype processes for the DM production can be summarized either by 
a quartic interaction like $\lambda \, \zeta_{3} \zeta_{4} J_{\rm DM}^2 $ and 
${\displaystyle \frac{\lambda^{\prime}}{v_S}} \, \overline{\mathcal{N}}_1 \mathcal{N}_2 J_{\rm DM}^2$, 
or a triple interaction like $ y \, \overline{\mathcal{N}}_1 \mathcal{N}_2 J_{\rm DM}$ 
and $\lambda^{''} v_S \, \zeta_{3} \zeta_{1} J_{\rm DM} $.

Focusing now only on the couplings of the scalar sector, we realize that the ones 
between $J_{\rm DM}$--$\zeta_{3}$--$\zeta_{4}$, and $\zeta_{3}$--$\zeta_{4}$--$\zeta_{5}$ 
(and combinations) are not necessarily suppressed, and they are controlled mainly 
by $\lambda_{X,S}$. Oppositely, the coupling in $J_{\rm DM}$--$\zeta_{5}$ is suppressed 
by the ratio $\left(M_J/v_S\right)^2$. It is worth mentioning that since the couplings 
controlling interactions with SM particles are mainly supressed, there is no thermal 
equilibrium between the Dark Sector (namely, $\zeta_1$ and $\zeta_2$) and the SM model particles, which makes the field 
$\zeta_1$ decoupled from the visible sector as well. Therefore, there is no intrusion 
of this field in the number of effective neutrino degrees of freedom~\cite{Weinberg:2013kea,Husdal:2016haj}.

Moving on to the fermion couplings for $J_{DM}$, the lagrangian $\mathcal{L}_{\rm int} = y_L \bar{L} H N_1$, 
gives rise to the interactions $\nu \zeta_5 \mathcal{N}_{1,2}$ with the yukawa coupling 
being proportional to $m_D$. Besides, the lagrangian $\mathcal{L}_{\rm int} = y_S S N_1 N_2$, 
where the coupling is proportional to $M$, gives rise to interactions $J_{\rm DM} \mathcal{N}_{1,2} \mathcal{N}_{1,2}$.
In both cases, the couplings are controlled by the inverse seesaw and, in our setup, they 
are around $y_L \simeq 0.1$ and $y_S < 10^{-3}$. These interactions produce an interplay 
between neutrinos, Higgs, and DM, in a similar fashion as in the scalar sector between 
light and heavy states. Since the interactions with light states are inverse-seesaw-supressed 
and heavy particles will decay, the thermalization is difficult to achieve for $J_{DM}$
by means of neutrino-driven processes.

Therefore, in the Early Universe, the evolution of the DM yield depends directly on the interaction of $\zeta_{3,4}$ and/or $\mathcal{N}_{1,2}$ with the SM.
In this way, the yields of LOSPs act as portals between SM and DM.
The combined processes are present whereas $T \gtrsim m_{\rm LOSP}$.
This means that LOSPs are likely in thermal equilibrium with the visible sector (i.e. the SM).
After that, they will decouple from the thermal bath in a similar way to the freeze-out, transferring subsequently their yields to $J_{\rm DM}$ and $\zeta_1$ via LOSP decays~\cite{Frigerio:2011in}.
However, not all of the final DM yield comes necessarily from these decays.
If the couplings among $J_{DM}$ and the LOSP are large enough, a portion of
DM yield might be reached being assisted by LOSP's interactions with the visible 
sector \emph{a-la} Freeze-out, until the LOSP is decoupled.

In the case of small $J_{DM}$-LOSP couplings, the outcoming fraction of the 
DM could be explained via freeze-in. A complete calculation of the DM abundance 
is given in Ref.~\cite{Boulebnane:2017fxw} where they consider the inverse 
seesaw scheme including a massive Majoron. They found that for the $v_{S}$ scale 
considered here, the DM production occurs above electroweak scale i.e. $m_{\mathcal{N}_i} >$~TeV 
with the leading process being $\mathcal{N}_{2} \rightarrow \mathcal{N}_{1} J_{\rm DM}$.
However, a full calculation for the DM relic abundance within this context is a 
matter of future work.

\section{\label{sec:disc} Discussion}

In the previous sections, we have described $J_{\rm DM}$ decay into neutrinos and three $\zeta_1$. In this section, we aim to perform a numerical analysis exploring
the stability of the DM candidate. As it was advanced, this is the observable that mostly constraints the model setup. Since the DM lifetime is expected to be extremely large with no stabilizing mechanism, it is foreseen that the correlations among some of the parameters must be strong, as it was advanced in the section~\ref{sec:decay}. 
Those correlations might hint an underlying unified symmetry, although we won't concern about this point. In this part, however, constraints coming from the DM relic abundance are not going to be considered.

\begin{table}[b]
\centering
\begin{tabular}{|c|c|}
\hline
Parameter & Value \\
\hline
\hline
$M$       & 100 TeV  \\
$\mu$     & 10 MeV  \\
$m_D$     & 10 GeV  \\
\hline
$v_S$     & $10^3$ -- $10^{8}$ GeV\\
$\omega$  & 0.4 -- 1.6 \\
\hline
\end{tabular}
\caption{
\label{tab:fixedpar}
Benchmarks and scan range for parameters in the $J_{\rm DM} \rightarrow \nu \nu$ decay.
}
\end{table}

%
In the first place, the channel to neutrinos will be analyzed. A first matter to notice concerns the couplings $O_L$ and $O_R$ shown at the Eq.~\ref{eq:nudecay}. In fact, they depend on the parameters $m_D$, $M$, $\omega$, $v_S$, and $m_{\nu}$. The latter is just a convenient parametrization in our setup with one massive neutrino, which has been made compatible with $m_\nu\, \sim 0.1$~eV (See Tab.~\ref{tab:fixedpar}). A more careful consideration on it should not affect the results in this section. However, as it was advanced in the section \ref{sec:decay}, we can set $\omega$ around $\sqrt{2/3}$ while fulfilling Eq.~\ref{eq:omega23}, so that the decay $J_{\rm DM}$ to neutrinos is produced in cosmological times by suppressing $O_L$ and $O_R$ simultaneously.

\begin{figure}[t]
\centering
\includegraphics[width=0.7\textwidth]{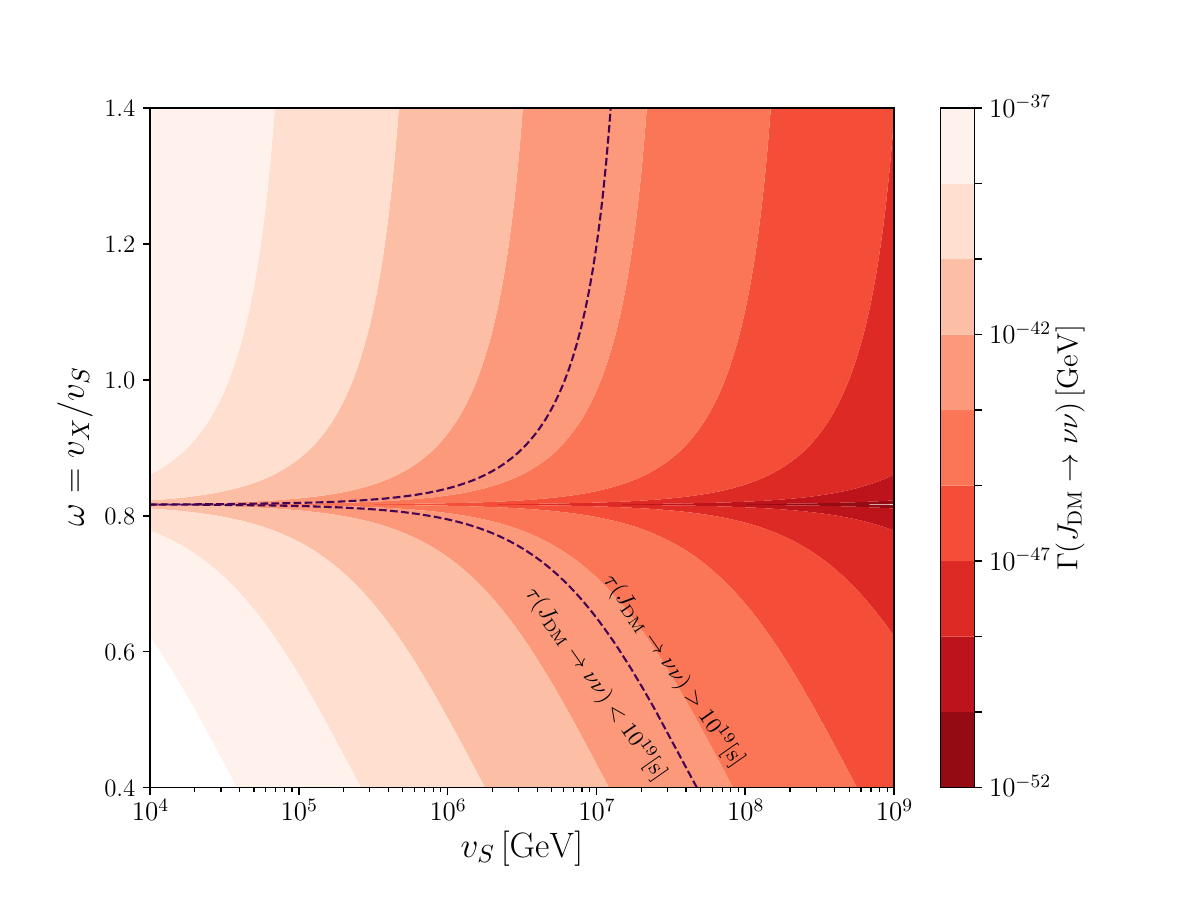}
\caption{\label{fig:omegavs} Plot $v_S$ versus $\omega$. The color palette indicates the value of the $J_{\rm DM}$ decay width to neutrinos. The dashed line shows the 
benchmark value for the DM lifetime $\tau_{DM} = 10^{19}$~s. The value $\omega = \sqrt{2/3}$ makes the decay width vanish regardless of $v_S$ value. For $v_S \gg 10^{6}$~GeV, $\omega$ starts to be irrelevant to satisfy the DM lifetime constraint.}
\end{figure}

In Fig.~\ref{fig:omegavs}, we present the result of a scan on the $J_{\rm DM}$ decay width in the plane $v_S$ versus $\omega$. It is shown, however, the way the decay width varies when having $\omega$ ranging between the values shown at Tab.~\ref{tab:fixedpar}. The dashed line on the plot shows the frontier of the DM lifetime given in seconds. For values of $v_S$ smaller than $10^6$ GeV, the DM lifetime requires that omega must be very close to $\sqrt{2/3}$, indicating a rather strong vev alignment among $S$ and $X$. On the contrary, larger values of $v_S$ weaken this alignment, since there is an additional suppression for the decay given by an overall factor $v_S^{-2}$ in the decay width (Eq.~\ref{eq:gammanu}).

On the other hand, the decay in the scalar sector can be analyzed separately since there is little intertwining between this decay channel and the one going to neutrinos, and it is given only by $\omega$ and $v_S$. As it was described in the previous section, the approach is to get rid of the unsuppressed contribution coming from $\lambda_{2111}\,-\,\frac{\lambda_{213}\lambda_{113}}{m_3^2}\,-\,\frac{\lambda_{214}\lambda_{114}}{m_4^2}$,  and take advantage of the suppression $(M_J/vs)^2$ arising as a residual. The Higgs-driven part of the decay has been neglected (see Eq.~\ref{eq:higgsmed}). Thus, we look for an interplay among the parameters $A$, $\psi$, $\lambda_X$, and $\omega$ that makes $\lambda^{\rm eff}_{2111}$ at Eq.~\ref{eq:scaleff} $\sim (M_J/v_S)^2$.

\begin{figure}[t]
\centering
\includegraphics[width=0.75\textwidth]{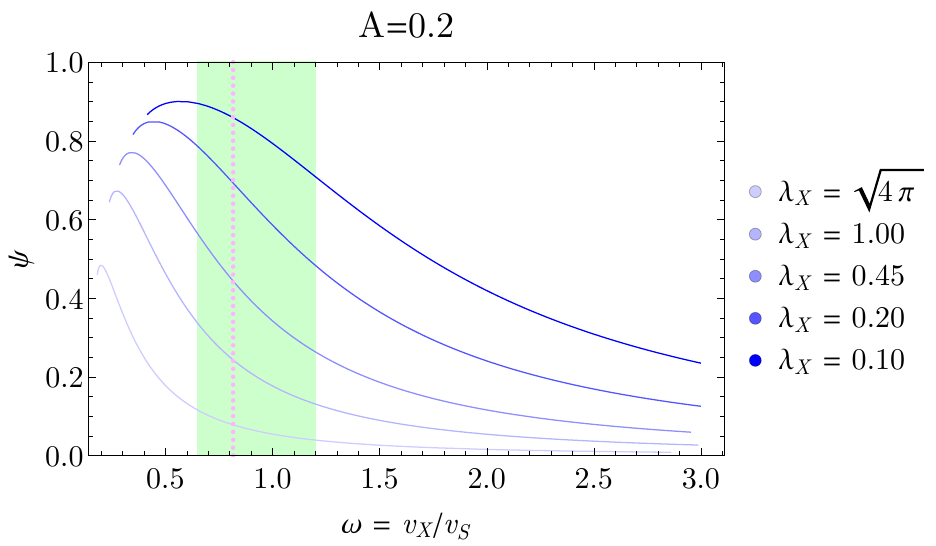}
\includegraphics[width=0.75\textwidth]{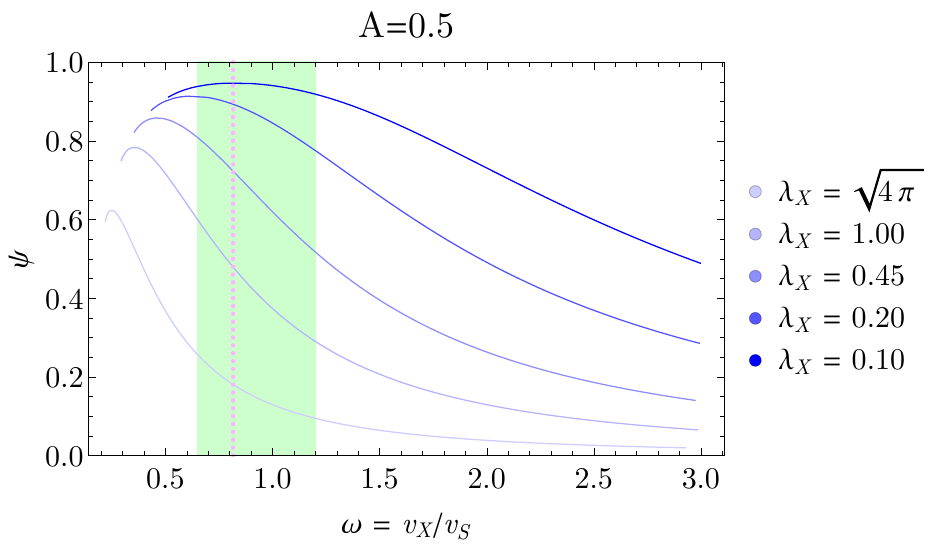}
\caption{\label{fig:psivsomega} Plot $\psi$ versus $\omega$ for $A=0.2$ (top) and $A=0.5$ (bottom).
The bluish lines correspond for the combination of $\psi$ and $\omega$ for a fixed value of 
$\lambda_X$ that makes the decay $J_{\rm DM}\rightarrow 3\zeta_1$ zero. The vertical magenta 
dashed line correspond to $\omega = \sqrt{2/3}$. The green area is the $\omega$ range that 
passes the DM lifetime constraint for the neutrino channel for $v_S \simeq 10^{6}$~GeV.}
\end{figure}

In Fig.~\ref{fig:psivsomega}, we show the combinations of $\psi$ and $\omega$ that suppress the decay width $J\longrightarrow 3\zeta_1$ to $\sim (M_J/v_S)^4$ for two values of A, $(0.2, 0.5)$ and for five values of $\lambda_X$ in different tones of blue. Notice that $\omega$ has been left as a free parameter. This selection was made in order to show a general trend in the dependence between $\psi$ and $\omega$. The blue curves range from the largest possible $\lambda_X$ (lightest blue line) given by the perturbative limit, to a smaller value ($\lambda_X = 0.1$, darkest blue) which has been chosen just in order to spot the trend. The leftmost value of each curve indicates the minimal solution for $\omega$ so that $\psi(\omega)$ starts becoming complex or stops representing a cosine of an angle. The light green zone corresponds to the range of $\omega$ compatible with the decay to neutrinos for $v_S \sim 10^{6}$~GeV, while the vertical dashed line is simply $\omega = \sqrt{2/3}$.

The scanned range of $\omega$ is made up to $3$ in order to explore a space with $v_X$ and $v_S$ in the same order of magnitude. However, this is not mandatory although it leads to other ways to control the DM half-life, other than the ones presented in previous section. Moreover, we have chosen our parameter space so that the heaviest scalars are above the TeV scale. By comparing both plots in Fig.~\ref{fig:psivsomega}, we observe that the perturbative limit for $\lambda_X$ sets a minimum $\psi(\omega)$-curve. That curve moves upwards insofar the value of $A$ is increased. On the contrary, the maximum $\psi(\omega)$-curves are related with the smallness of $\lambda_X$, however $\lambda_X \simeq 0$ points towards an unstable vacuum. 

\begin{figure}[t]
\includegraphics[width=0.75\columnwidth]{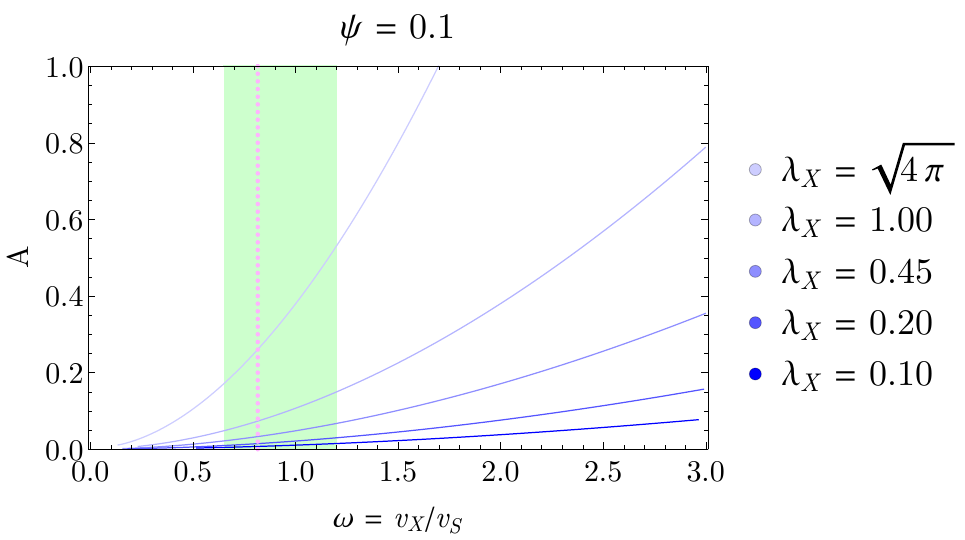}
\includegraphics[width=0.75\columnwidth]{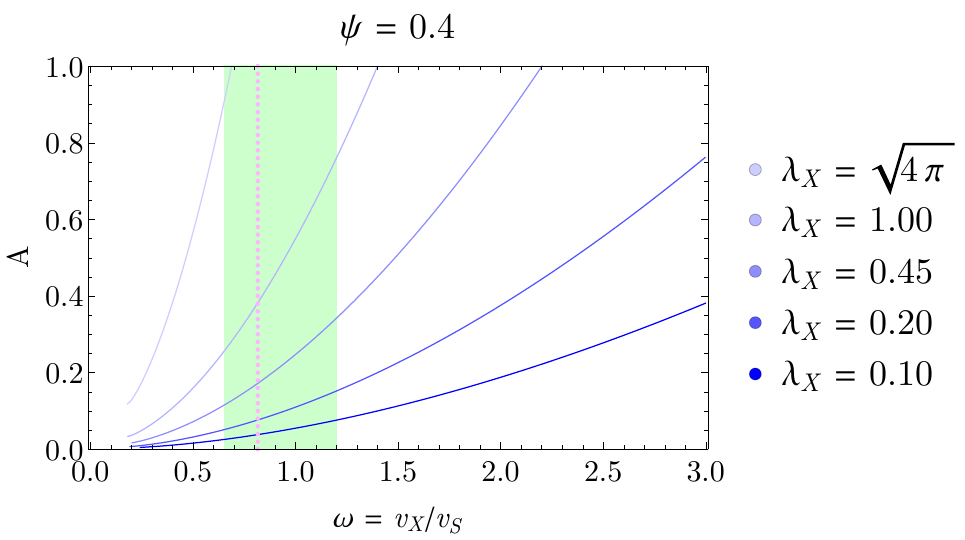}
\centering
\caption{
\label{fig:Avsomega}
Plot $A$ versus $\omega$ for $\psi=0.1$ (top) and $\psi=0.4$ (bottom). The bluish lines correspond for the combination of $A$ and $\omega$ for a fixed value of $\lambda_X$ that makes the decay $J_{\rm DM}\rightarrow 3\zeta_1$ zero. The vertical magenta dashed line correspond to $\omega = \sqrt{2/3}$. The green area is the $\omega$ range that passes the DM lifetime constraint for the neutrino channel for $v_S \simeq 10^{6}$~GeV.
}
\end{figure}

Similar information is shown in Fig.~\ref{fig:Avsomega}, where we present the suppressed decay width solution for $A(\omega)$. Here we observe that the pertubative limit of $\lambda_X$ produces an upper-bound-$A(\omega)$-curve for each choice of $\psi$. For both Figs.~\ref{fig:psivsomega} and~\ref{fig:Avsomega}, we show that there is a smooth transition for different values of $\lambda_X$ and the combinations of $\psi$, $A$, and $\omega$ that make the decay width $\sim (M_J/v_S)^4$. This implies that the solutions belong to a smooth volume in the parameter space, and therefore, one can always find one parameter when the other three have been given. Besides, we find that extreme values of $A \, (\sim 0)$, and $\psi \, (\gtrsim 1.0 \, , \lesssim   0.0)$, are not favored by the DM stability condition and these values could lead to tachyonic states of $\zeta_3$ or $\zeta_4$. Moreover, when we focus on the green region, which was indicated as the zone in $\omega$ for which the decays to neutrinos are suppressed, we find that the most of the curves pass through it. This is showing that there is a natural compatibility among the solutions for the neutrino and scalar decay modes independently.

An interesting case involves the Higgs physics. In our model, the mixing among the scalars gives rise to a SM-like Higgs that is weakly mixed with the rest of the scalars by a factor $\epsilon_h \sim v_h/v_S$. However, this mixing does not forbid a sizeable contribution to the invisible Higgs decay, namely $H \rightarrow J_{\rm DM} J_{\rm DM},\, J_{\rm DM} \zeta_1,\, \zeta_1 \zeta_1$. These processes come directly from the scalar potential via $\lambda_{215}$, $\lambda_{115}$, and $\lambda_{225}$ (See Eqs.~\ref{eq:l215}, \ref{eq:l115}, and \ref{eq:l225}, respectively). We observe that all these couplings are suppressed by $(M_J/v_S)^2$, therefore the decay width for these channels is suppressed by $(M_J/v_S)^4$, i.e. a suppression compatible to that for $J_{DM}$ lifetime. After evaluation, we obtain that the Higgs decay width to invisibles in our scenario is $\mathcal{O}\left( 10^{-44} \right)$~GeV, and thus these processes cannot be constrained using the measurement of the invisible higgs decay~\cite{Chatrchyan:2014tja,Baek:2014jga} since they are far below the upper limit.

The role of CP-phases in the decay width either in the scalar or neutrino modes is not an issue. In the scalar sector, the effect is washed out by the tadpole equations that fix the relation among the 3 phases: $\theta$, $\tau$, and $\delta$. In the case of neutrinos, in addition to those CP-phases, we could include extra phases in the yuwakas: $y_L$, $y_S$, and $y_X$, which are $3\times1$, $1\times1$ and $1\times1$ matrices in our setup. However, we decided to keep the inverse seesaw mass terms real, and hence, the possible impact of CP-phases in the phenomenology is absorbed, as it is shown throughout section~\ref{sec:model}, in particular in eqs.~\ref{eq:gamma1},~\ref{eq:gamma2} and~\ref{eq:gamma3}. If we wanted to add effect of CP-phases, we should either relax the condition of real mass terms or add more families of neutrinos.

This addition of CP-phases effects in the DM decay adds an improvement on this setup. A different improvement is to promote from a global $U(1)_l$ symmetry to a gauge one. If that symmetry is anomalous given the charge assignments made in our model, is not an issue either. A heavy field, charged under that group, carrying the opposite anomalous numbers can be always included, and since it is heavy, its effect on the dynamics might be integrated out. This would relax the correlations in the scalar sector, because the $\zeta_1$ would be absorbed as the corresponding longitudinal degree of freedom for the $U(1)_l$ gauge boson after the SSB, and it won't be a physical degree of freedom anymore. The latter feature is going to be exploited in a future work.

\section{\label{sec:conc}Conclusions}

In this work we proposed an extension of SM where neutrinos become massive through an inverse seesaw mechanism arising after spontaneous breaking of the lepton number and electroweak symmetries. As a product of this breaking, we have a pseudoscalar sector that has: A $SU(2)_L\times U(1)_Y$ Nambu-Goldstone boson which is disconnected from other pseudoscalars, a massless pseudoscalar coming after $U(1)_l$ breaking (dubbed as \emph{massless Majoron}), and a massive pseudoscalar which we have named as \emph{massive Majoron}. On the other hand, the scalar sector contains a light scalar that has been identified with the Higgs boson, and two scalars that correspond to \emph{heavy scalars}. From the former sector, the massive Majoron has been identified as a potential DM candidate, although it decays mainly into neutrinos and massless Majorons.

To stabilize the lifetime of the massive Majoron with the addition of no \emph{ad-hok} discrete symmetries, we have chosen to have a DM candidate with masses around keV.
This mass range choice reduces the decay channel into massless Majorons in a straightforward way. If we choose to have a heavier DM candidate, we should proceed similarly, although through this work we remain in the limit of a keV Majoron. 

The introduced scalars which give rise to the inverse seesaw mechanism, also allow the spontaneous breaking of CP invariance. Nevertheless, the effect is not present in phenomenology of our model because we included just one family of active neutrinos.

The DM candidate stability is very fragile in this model because we did not include any \emph{ad-hoc} stabilizing symmetry. However, we found that there is always a region in the parameter space where the massive Majoron has a lifetime longer than $10^{19}$~s and, therefore, it can be considered as a plausible DM candidate.

Moreover, we found that the ratio among vevs, $\omega$, has a very important role in the decay channel to neutrinos. The value $\omega = \sqrt{2/3}$ can vanish the decay mode into neutrinos presenting a tantalizing vev alignment for model building. The scalar decay modes are the most crucial because the drastic effect on the total DM lifetime and from the point of view of scan of the parameter space. Nevertheless, we found that the decay width vanished in a region of the parameter space of the scalar sector. We also discussed how to rid off the scalar decay modes by promoting the global lepton symmetry to a local one. We discuss possible ways on how to estimate the DM relic abundance in terms of a freeze-in scenario.

In general words, the model presents an interesting relation between the neutrino mass mechanism and origin of the massive Majoron as a DM candidate.

\appendix
\section{Lepton charge assignments} \label{app:charge}
In this section, we examine the most general way in which the lepton charges are fixed for the fields $N_1$, $N_2$, $S$ and $X$.
As we advanced, the Yukawa couplings at Eq.~\ref{eq:slangran} will fix the value for the lepton numbers of the field $N_1$. However, this does not fix the charges for the new scalars $S$ and $X$, nor for $N_2$.
The final assignation can be obtained after considering the following general scalar potential:
\begin{eq}
V_{\rm I} = \lambda_{J} \, e^{i\delta} X^{m}S^{\dagger n} + {\rm h.c.} \, , \label{eq:deltagen}
\end{eq}

After demanding that $V_I$ is renormalizable, we can choose the values of $m$ and $n$ that subsequently will fix the values of lepton number for $S$ and $X$.
Thus, one has a collection of models formed by taking $m+n=2,3,4$.
Notice that we still have to choose one value for $m$ and $n$.
By now, we choose $m+n=4$ and, by following the assignation made at the Table~\ref{tab:charges}, we establish conditions in order
to make \ref{eq:deltagen} invariant under lepton number.
\begin{eq}
m\,+\,n\, =\, 4 \quad {\rm and}\quad m\left( 2x\right)\,-n\left( 1\,-\,x\right)\, =\, 0 \quad \Rightarrow \nonumber \\
x\, =\, \frac{n}{n+2m}\, =\, \frac{n}{8-n}
\end{eq}

Recall that $n,m$ are integers running from $1\dots 3$ ($0$ and $4$ will break lepton number explicitly).
Therefore, for $n=3$ and $m=1$, one has $x=3/5$, as it was stated in the section~\ref{sec:spinvseesaw}.
At the Table~\ref{tab:chgen}, we present the lepton number charges for different values of $n$ for $m+n=4$.
In order to show a case with a different choice of $m+n$, at the Table~\ref{tab:chgen2} we present the lepton numbers of the fields after considering $m+n=3$ at Eq.~\ref{eq:deltagen}.

\begin{table}[tb]
\centering
\begin{tabular}{|c|ccccc|}
\hline
        &  $L$  &  $N_1$  &  $N_2$  &  $S$  &  $X$  \\
\hline\hline
 $n=1$  & $1$   & $-1$ & $1/7$  & $6/7$   & $2/7$ \\
 $n=2$  & $1$   & $-1$ & $1/3$  & $2/3$   & $2/3$ \\
 $n=3$  & $1$   & $-1$ & $3/5$  & $2/5$   & $6/5$ \\
\hline
\end{tabular}
\caption{\label{tab:chgen} Charge assignment of different models for $m+n=4$.}
\end{table}

\begin{table}[tb]
\centering
\begin{tabular}{|c|ccccc|}
\hline
        & $L$  &  $N_1$  &  $N_2$  &  $S$  &  $X$  \\
\hline\hline
 $n=1$  & $1$   & $-1$ & $1/5$  & $4/5$   & $2/5$ \\
 $n=2$  & $1$   & $-1$ & $1/2$  & $1/2$   & $1$ \\
\hline
\end{tabular}
\caption{\label{tab:chgen2} Charge assignment of different models for $m+n=3$.}
\end{table}

\section{Couplings}

In this section we describe briefly the relevant couplings used in this work, with an special emphasis in the interactions participating in
the decays of the Majoron.

\subsection{Fermion Couplings} \label{app:fercoup}
The couplings shown below are related to the process $J_{DM} \longrightarrow \nu\nu$ that appears in the models involving majoron DM. Recall that in these couplings we got rid of the explicit dependence of the Yukawas and it was preferred to work with the mass parameters involved in neutrino mass generation via inverse seesaw, namely $\mu$, $m_D$ and $M$ (c.f. Eq.~\ref{eq:nudecay})

\begin{eq}
O_L &=& \frac{\mathcal{D}_1^{(L)}}{\mathcal{D}_2^{(L)}} \\
\mathcal{D}_1^{(L)} &=& im_\nu \left(4 m_D^6 + 4 M m_D^4 m_\nu +
   M^3 m_\nu^3\right) \nonumber \\
   & & \left[ \left( -4 m_D^8 + 4 M^2 m_D^4 m_\nu^2 -
      M^3 m_D^2 m_\nu^3 + M^4 m_\nu^4\right) \right. \nonumber \\
    &+& \left. 3 \left(2 m_D^8 + 2 M m_D^6 m_\nu +
      M^4 m_\nu^4\right) \omega^2 \right] \\
\mathcal{D}_2^{(L)} &=& \left(2m_D^2 + 3 M m_\nu\right)^2 \left(m_D^2 -
   M m_\nu\right) \nonumber \\
   & & \left(2 m_D^4 - M m_D^2 m_\nu + M^2 m_\nu^2\right)^2 v_s \omega \sqrt{2 + 18 \omega^2} \\
O_R &=& \left( O_L \right)^{*}\, .
\end{eq}

By using the definitions from above, the function $f$ at Eq.~\ref{eq:J2nu} can be expressed as
\begin{eq}
f\left( m_\nu,m_D,M,v_S \right) = \left|\left| O_L \right|\right|^2\,+\, \left|\left| O_R \right|\right|^2
\end{eq}

\subsection{Scalar Couplings} \label{app:scacoup}

Since the relevant couplings for the DM decay in the scalar sector are cuartic, they have no mass dimensions.
This is respected by the effective coupling at Eq.~\ref{eq:scaleff}, which makes the entire coupling independent of mass scales, and thus, it just depends on the adimensional parameters we set (namely $A$, $\omega$, $\psi$, $\lambda_h$, $\lambda_X$ and $M_J/v_S$).

First, it is shown the formula for the direct contribution to this coupling:
\begin{eq}
\lambda_{2111} &=& \frac{\mathcal{D}_1^{(1)}}{\mathcal{D}_2^{(1)}} \\
\mathcal{D}_1^{(1)} &=& -3 \left[3 A \left(1 + 9 \omega^2\right) \left\{-2 \psi \omega +
       \sqrt{1 - \psi^2} \left(-1 + 9 \omega^2\right)\right\}  \right. \nonumber \\
       &+&  \left.  8 \psi \omega \left\{ \left(\frac{M_J^2}{v_s^2}\right) -
       27 \left( \frac{M_J^2}{v_s^2} \right) \omega^2 +
       6 \lambda_{X} \omega^2 \left(1 + 9 \omega^2\right)\right\}\right] \\
\mathcal{D}_2^{(1)} &=& 4 \psi (1 + 9 \omega^2)^3
\end{eq}

Now, we show the formuli for the contributions coming from the integrated effect of the heavy scalars.
On the one hand, we explicit the formuli for $\frac{\lambda_{213}\lambda_{113}}{m_3^2}$ and $\frac{\lambda_{214}\lambda_{114}}{m_4^2}$,
which share some similarities.

\begin{eq}
\label{eq:lam213}
\frac{\lambda_{213}\lambda_{113}}{m_3^2} &=& \frac{\mathcal{D}_1^{(3)}}{\mathcal{D}_2^{(3)}} \\
\mathcal{D}_1^{(3)} &=&  3 \left[A \left(-1 + \psi\right) \left(-1 + \psi -
        9 \sqrt{1 - \psi^2} \omega\right) \left(1 + 9 \omega^2\right) \right.  \nonumber \\
          &-&  \left. 2 \psi \omega \left(12 \left(\frac{M_J^2}{v_s^2} \right) \left(\sqrt{1 - \psi^2} +
           5 \left(-1 + \psi\right) \omega\right)  \right. \right. \nonumber \\
          &+& \left. \left. \lambda_{X} \omega \left(1 - \psi + 9 \sqrt{1 - \psi^2} \omega\right) \left(1 +
           9 \omega^2\right)\right)\right]  \nonumber \\
           & &  \left[A \left(-1 + \psi\right) \left(\sqrt{1 - \psi^2} + \left(-1 + \psi\right) \omega\right) \left(1 +
        9 \omega^2\right) \right.  \nonumber \\
           &+&   \left.  2 \psi \omega \left(\lambda_{X} \omega \left(\sqrt{1 - \psi^2} + \left(-1 + \psi\right) \omega\right) \left(1 +
           9 \omega^2\right) \nonumber \right. \right.  \\
           &-&   \left. \left. 4 \left(\frac{M_J^2}{v_s^2} \right) \left(1 + 3 \sqrt{1 - \psi^2} \omega -
           6 \omega^2 + \psi \left(-1 +
              6 \omega^2\right)\right)\right)\right] \\
\mathcal{D}_2^{(3)} &=& 8 \left(-1 + \psi\right) \psi \left(1 + 9 \omega^2\right)^3 \nonumber \\
                    & & \left[-A \left(1 + \psi\right) \left(1 + 9 \omega^2\right) +
     2 \psi \left(-\lambda_{X} \omega^2 \left(1 +
           9 \omega^2\right) \right. \right. \nonumber \\
           &+& \left. \left. \left( \frac{M_J^2}{v_s^2} \right) \left(-1 + \psi -
                6 \sqrt{1 - \psi^2} \omega + 3 \omega^2 +
                3 \psi \omega^2\right)\right)\right]\, ,
\end{eq}

\begin{eq}
\label{eq:lam214}
\frac{\lambda_{214}\lambda_{114}}{m_4^2} &=& \frac{\mathcal{D}_1^{(4)}}{\mathcal{D}_2^{(4)}} \\
\mathcal{D}_1^{(4)} &=& 3 \left[ A \left( 1 + \psi\right) \left( 1 + \psi -
        9 \sqrt{1 - \psi^2} \omega\right) \left( 1 + 9 \omega^2\right) \nonumber \right. \nonumber \\
                    &-& \left. 2 \psi \omega \left( 12 \left( \frac{M_J^2}{v_s^2} \right) \left( \sqrt{
           1 - \psi^2} +
           5 \left( 1 + \psi\right) \omega\right) \right. \right. \nonumber \\
           &-&  \left. \left. \lambda_{X} \omega \left( 1 +
\psi - 9 \sqrt{1 - \psi^2} \omega\right) \left( 1 +
           9 \omega^2\right)\right)\right] \nonumber \\
           & & \left[ A \left( 1 + \psi\right) \left( \sqrt{
        1 - \psi^2} + \omega + \psi \omega\right) \left( 1 +
        9 \omega^2\right) \right. \nonumber \\
        &+&  \left. 2 \psi \omega \left( \lambda_{X} \omega \left( \sqrt{
           1 - \psi^2} + \omega + \psi \omega\right) \left( 1 +
           9 \omega^2\right) \right. \right. \nonumber \\
         &-&  \left. \left. 4 \left( \frac{M_J^2}{v_s^2} \right) \left( -1 +
           3 \sqrt{1 - \psi^2} \omega +
           6 \omega^2 + \psi \left( -1 +
              6 \omega^2\right)\right)\right)\right] \\
\mathcal{D}_2^{(4)} &=& 8 \psi \left(1 + \psi\right) \left(1 + 9 \omega^2\right)^3 \nonumber \\
                   & & \left[A \left(-1 + \psi\right) \left(1 + 9 \omega^2\right) +
     2 \psi \left(\lambda_{X} \omega^2 \left(1 +  9 \omega^2\right) \right. \right. \nonumber \\
                   &+& \left. \left. \left(\frac{M_J^2}{v_s^2} \right) \left(1 + \psi -
           6 \sqrt{1 - \psi^2} \omega - 3 \omega^2 +
           3 \psi \omega^2\right)\right)\right]
\end{eq}

The contributions of the Higgs field to the majoron decay (the fraction $\frac{\lambda_{215}\lambda_{115}}{m_5^2}$) are written below.
The  expressions for $\lambda_{215}$, $\lambda_{225}$ and $\lambda_{115}$ are proportional to the ratio $\left( \frac{M_J}{v_s} \right)^4$, thus, having a keV majoron implies a natural supression for the contributions to the decay.

\begin{eq}
\frac{\lambda_{215}\lambda_{115}}{m_5^2} &=& \frac{\mathcal{D}_1^{(5)}}{\mathcal{D}_2^{(5)}} \\
\mathcal{D}_1^{(5)} &=& -1152 \left( \frac{M_J}{v_S} \right)^4 \psi^2 \omega^3 \left[2 \psi \left(5
\lambda_{HS} - \lambda_{HX}\right) \lambda_{X} \omega^3 \right. \nonumber \\
   &-& \left.  A \left \{\sqrt{1 - \psi^2} \lambda_{HS} + \lambda_{HX} \omega \left(2
\psi - 5 \sqrt{1 - \psi^2} \omega\right)\right\}\right] \nonumber \\
    & & \left[-2 \psi \lambda_{X} \omega \left(\lambda_{HS} - 6 \lambda_{HS} \omega^2 +
         3 \lambda_{HX} \omega^2\right) \right. \nonumber \\
    &-& \left. A \left\{3 \sqrt{1 - \psi^2} \lambda_{HS} + \lambda_{HX} \left(6 \psi \omega + \sqrt{1 - \psi^2}
      \left(1 - 6 \omega^2\right)\right)\right\}\right]\nonumber \\ \\
\mathcal{D}_2^{(5)} &=& \left(1 + 9 \omega^2\right)^3 \nonumber \\
               & & \left[-A^4 \left(-1 + \psi^2\right)^2 \lambda_{h} \left(1 + 9 \omega^2\right) +
      8 A^3 \psi \left(-1 + \psi^2\right) \omega \right. \nonumber \\
               & & \left. \left(2 \sqrt{1 - \psi^2} \lambda_{HS} \lambda_{HX} + \psi \left(2
\lambda_{HX}^2 - \lambda_{h} \lambda_{X}\right) \omega\right) \left(1 + 9 \omega^2\right) \right. \nonumber \\
               &+& \left. 16 \psi^4 \lambda_{X}^2 \omega^6 \left[8 \left(\frac{M_J}{v_S}\right)^2
\left(3 \lambda_{HS}^2 +  6 \lambda_{HS} \lambda_{HX} - \lambda_{HX}^2\right) \right. \right. \nonumber \\
               &+& \left. \left. \lambda_{X}
\left(1 + 9 \omega^2\right) \left(4 \lambda_{HS}^2 + \left(4 \lambda_{HX}^2 - \lambda_{h}
\lambda_{X}\right) \omega^2\right)\right] \right. \nonumber \\
               &+&  \left. 32 A \psi^3 \lambda_{X} \omega^3 \left[\lambda_{X} \omega \left(1 +
            9 \omega^2\right) \left\{2 \sqrt{1 - \psi^2} \lambda_{HS} \lambda_{HX} \omega \right. \right. \right. \nonumber \\
               &+& \left. \left. \left. \psi \left(2 \lambda_{HS}^2 + \left(4 \lambda_{HX}^2 - \lambda_{h} \lambda_{X}\right)
\omega^2\right)\right\}  \right.  \right. \nonumber \\
               &+& \left. \left. 4 \left(\frac{M_J}{v_S}\right)^2 \left\{ 3 \sqrt{1 - \psi^2} \lambda_{HS}^2 + \lambda_{HX}^2 \omega
\left(-2 \psi + 3 \sqrt{1 - \psi^2} \omega\right) \right. \right. \right. \nonumber \\
               &+& \left. \left. \left. \lambda_{HS} \lambda_{HX} \left(6 \psi \omega + \sqrt{1 - \psi^2}
                                             \left(-1 + 3 \omega^2\right)\right)\right\}\right] \right. \nonumber \\
               &+& \left. 8 A^2 \psi^2 \left\{\lambda_{X} \omega^2 \left(1 + 9 \omega^2\right)
                                              \left(2 \left(-1 + \psi^2\right) \lambda_{HS}^2 \right. \right. \right. \nonumber \\
               &+& \left. \left. \left. 8 \psi \sqrt{1 - \psi^2} \lambda_{HS} \lambda_{HX} \omega + \left(2
\left(-1 + 5 \psi^2\right) \lambda_{HX}^2 + \left(1 - 3 \psi^2\right) \lambda_{h} \lambda_{X}\right) \omega^2\right) \right. \right. \nonumber \\
               &+& \left. \left. 4 \left(\frac{M_J}{v_S}\right)^2 \left[\left(-1 + \psi^2\right) \lambda_{HS}^2 -
         2 \lambda_{HS} \lambda_{HX} \omega \left(2 \psi \sqrt{1 - \psi^2} - 3 \omega +
               3 \psi^2 \omega\right) \right. \right. \right. \nonumber \\
               &+& \left. \left. \left. \lambda_{HX}^2 \omega^2 \left(12 \psi \sqrt{1 - \psi^2} \omega
                    + 3 \omega^2 - \psi^2 \left(4 + 3 \omega^2\right)\right)\right]\right\}\right]
\end{eq}

Finally, we show the expressions for the couplings that lead to the Higgs invisible decays $H\longrightarrow 2 \zeta_1$, $H\longrightarrow 2 \zeta_2$
and $H\longrightarrow \zeta_1 \zeta_2$.
These formuli show that the contributions from the new fields $\zeta_{1,2}$ to the invisible Higgs decay are heavily supressed, since they are proportional to $\left( \frac{M_J}{v_s} \right)^2$.
On top of that, observe that these expressions also depend on $\lambda_{HX}$ and $\lambda_{SH}$, couplings that have been taken to be $\ll 1$.

\begin{align}
\label{eq:l215}
\lambda_{215} &= -24 \left(\frac{M_J}{v_s}\right)^2 \, \frac{ \, v_h\, \psi\, \omega^2 }{ \left(1 + 9 \omega^2\right)^2 \left\{A^2 \left(-1 + \psi^2\right) +
   4 A \psi^2 \lambda_{X} \omega^2 +
   4 \psi^2 \lambda_{X}^2 \omega^4\right\} } \nonumber \\
              & \left[\lambda_{HS} \left\{ 3 A \sqrt{1 - \psi^2} +
    2 \psi \lambda_{X} \omega -
    12 \psi \lambda_{X} \omega^3\right\} + \lambda_{HX} \left\{6 \psi \lambda_{X} \omega^3 +
    A \left(6 \psi \omega +
       \sqrt{1 - \psi^2} \left(1 - 6 \omega^2\right)\right)\right\} \right]
\end{align}
\begin{align}
\label{eq:l115}
\lambda_{115} &= -24 \left(\frac{M_J}{v_s}\right)^2 \, \frac{ v_h\, \psi\, \omega}{ \left(1 + 9 \omega^2\right)^2 \left\{A^2 \left(-1 + \psi^2\right) +
   4 A \psi^2 \lambda_{X} \omega^2 +
   4 \psi^2 \lambda_{X}^2 \omega^4\right\} } \nonumber \\
       &  \left[ \lambda_{HS} \left\{ A \sqrt{1 - \psi^2} -
    10 \psi \lambda_{X} \omega^3\right\} + \lambda_{HX} \left\{2 \psi \
     \lambda_{X} \omega^3 +
    A \omega \left(2 \psi - 5 \sqrt{1 - \psi^2} \omega\right)\right\}\right]
\end{align}

\begin{align}
\label{eq:l225}
\lambda_{225} &= - 72 \left( \frac{M_J}{v_S}\right)^2 \frac{ v_h\, \psi\, \omega^3 }{\left(1 +
                      9 \omega^2\right)^2 \left\{A^2 \left(-1 + \psi^2\right) +
                      4 A \psi^2 \lambda_{X} \omega^2 +
                      4 \psi^2 \lambda_{X}^2 \omega^4\right\}} \nonumber \\
               &   \left[2 \psi \lambda_{X} \omega \left\{3 \lambda_{HX} \omega^2 + \lambda_{HS}
                  \left(2 + 3 \omega^2\right)\right\} + A \left\{3 \sqrt{1 - \psi^2} \lambda_{HS} + \lambda_{HX} \left(6 \psi
                  \omega + \sqrt{1 - \psi^2} \left(2 + 3 \omega^2\right)\right)\right\}\right]
\end{align}

\section{Two-Loop Running Couplings for the Scalar Sector} \label{sec:2loop}
{}In this appendix we show the running for the quartic couplings of the scalar
sector. Of a particular interest is the coupling $\lambda_{J}$, whose running is
proportional to itself at one and two-loops, which ensures the smallness of its
value at all orders in perturbation theory, and thus it ensures the smallness of the
mass of $J$. This should be present given the existence of an additional $U(1)$ symmetry
involving $S$ and $X$ even after the breakdown of electroweak symetry.
{}It is worthwhile to mention that the couplings $\lambda_{HX}$ and $\lambda_{HS}$
are not proportional to themselves, but they only depend on Yukawas giving rise to
the inverse seesaw parameters $m_D$ and $\mu$, and also on themselves. All of
them are small parameters that keep the interactions of $\zeta_5$ with $J$ almost
shut, which forbids overproduction of $J$ via Higgs decays.
The couplings $y_{(L,\,S,\,X)}$ are the Yukawa couplings defined in the Lagrangian
\ref{eq:slangran}. In turn, $y_{(u,\,d)}$ are Standard Model quark Yukawa couplings. Finally,
the quartic couplings $\lambda_{(H,\,5,\,S,\,X,\,HS,\,HX)}$ are defined at equations \ref{eq:potsx},
\ref{eq:potvi} and \ref{eq:higgsmix}. The calculation of these RGEs was performed by an implementation
of this model in SARAH~\cite{Staub:2013tta}.

\begin{align}
\beta_{\lambda_H}^{(1)} & =
+\frac{27}{100} g_{1}^{4} +\frac{9}{10} g_{1}^{2} g_{2}^{2} +\frac{9}{4} g_{2}^{4} -\frac{9}{5} g_{1}^{2} \lambda_H - 9 g_{2}^{2} \lambda_H +12 \lambda_H^{2} +2 \lambda_{HX}^{2} +2 \lambda_{HS}^{2} +12 \lambda_H \mbox{Tr}\Big({y_d  y_{d}^{\dagger}}\Big) \nonumber \\
 &  +4 \lambda_H \mbox{Tr}\Big({y_{e}  y_{e}^{\dagger}}\Big) +4 \lambda_H \mbox{Tr}\Big({y_{L}  y_{L}^{\dagger}}\Big) +12 \lambda_H \mbox{Tr}\Big({y_{u}  y_{u}^{\dagger}}\Big) -12 \mbox{Tr}\Big({y_d  y_{d}^{\dagger}  y_d  y_{d}^{\dagger}}\Big) \nonumber \\
 & -4 \mbox{Tr}\Big({y_{e}  y_{e}^{\dagger}  y_{e}  y_{e}^{\dagger}}\Big) - 4 \mbox{Tr}\Big({y_{L}  y_{L}^{\dagger}  y_{L}  y_{L}^{\dagger}}\Big) -12 \mbox{Tr}\Big({y_{u}  y_{u}^{\dagger}  y_{u}  y_{u}^{\dagger}}\Big)  \\
\beta_{\lambda_H}^{(2)} & =
-\frac{3411}{1000} g_{1}^{6} -\frac{1677}{200} g_{1}^{4} g_{2}^{2} -\frac{289}{40} g_{1}^{2} g_{2}^{4} +\frac{305}{8} g_{2}^{6} +\frac{1887}{200} g_{1}^{4} \lambda_H +\frac{117}{20} g_{1}^{2} g_{2}^{2} \lambda_H \nonumber \\
 &-\frac{73}{8} g_{2}^{4} \lambda_H +\frac{54}{5} g_{1}^{2} \lambda_H^{2} +54 g_{2}^{2} \lambda_H^{2} - 78 \lambda_H^{3} - 10 \lambda_H \lambda_{HX}^{2} -8 \lambda_{HX}^{3} -10 \lambda_H \lambda_{HS}^{2} -8 \lambda_{HS}^{3} \nonumber \\
 &+\frac{1}{10} \Big(225 g_{2}^{2} \lambda_H  -45 g_{2}^{4}  + 80 \Big(10 g_{3}^{2}  -9 \lambda_H \Big)\lambda_H + 9 g_{1}^{4}  + g_{1}^{2} \Big(25 \lambda_H  + 54 g_{2}^{2} \Big)\Big)\mbox{Tr}\Big({y_d  y_{d}^{\dagger}}\Big) \nonumber \\
 &-\frac{3}{10} \Big(15 g_{1}^{4}  + 5 \Big(16 \lambda_H^2 -5 g_{2}^{2} \lambda_H  + g_{2}^{4}\Big) - g_{1}^{2} \Big(22 g_{2}^{2}  + 25 \lambda_H \Big)\Big)\mbox{Tr}\Big({y_{e}  y_{e}^{\dagger}}\Big) \nonumber \\
 &-\frac{9}{50} g_{1}^{4} \mbox{Tr}\Big({y_{L}  y_{L}^{\dagger}}\Big) -\frac{3}{5} g_{1}^{2} g_{2}^{2} \mbox{Tr}\Big({y_{L}  y_{L}^{\dagger}}\Big) -\frac{3}{2} g_{2}^{4} \mbox{Tr}\Big({y_{L}  y_{L}^{\dagger}}\Big) +\frac{3}{2} g_{1}^{2} \lambda_H \mbox{Tr}\Big({y_{L}  y_{L}^{\dagger}}\Big) +\frac{15}{2} g_{2}^{2} \lambda_H \mbox{Tr}\Big({y_{L}  y_{L}^{\dagger}}\Big) \nonumber \\
 &-24 \lambda_H^{2} \mbox{Tr}\Big({y_{L}  y_{L}^{\dagger}}\Big) - \lambda_{HS}^{2} \mbox{Tr}\Big({y_{S}  y_{S}^{\dagger}}\Big) -\frac{171}{50} g_{1}^{4} \mbox{Tr}\Big({y_{u}  y_{u}^{\dagger}}\Big) \nonumber \\
 &+\frac{63}{5} g_{1}^{2} g_{2}^{2} \mbox{Tr}\Big({y_{u}  y_{u}^{\dagger}}\Big) -\frac{9}{2} g_{2}^{4} \mbox{Tr}\Big({y_{u}  y_{u}^{\dagger}}\Big) +\frac{17}{2} g_{1}^{2} \lambda_H \mbox{Tr}\Big({y_{u}  y_{u}^{\dagger}}\Big) +\frac{45}{2} g_{2}^{2} \lambda_H \mbox{Tr}\Big({y_{u}  y_{u}^{\dagger}}\Big) \nonumber \\
 &+80 g_{3}^{2} \lambda_H \mbox{Tr}\Big({y_{u}  y_{u}^{\dagger}}\Big) -72 \lambda_H^{2} \mbox{Tr}\Big({y_{u}  y_{u}^{\dagger}}\Big) -2 \lambda_{HX}^{2} \mbox{Tr}\Big({y_{X}  y_{X}^*}\Big) +\frac{8}{5} g_{1}^{2} \mbox{Tr}\Big({y_d  y_{d}^{\dagger}  y_d  y_{d}^{\dagger}}\Big) \nonumber \\
 &-64 g_{3}^{2} \mbox{Tr}\Big({y_d  y_{d}^{\dagger}  y_d  y_{d}^{\dagger}}\Big) -3 \lambda_H \mbox{Tr}\Big({y_d  y_{d}^{\dagger}  y_d  y_{d}^{\dagger}}\Big) -42 \lambda_H \mbox{Tr}\Big({y_d  y_{u}^{\dagger}  y_{u}  y_{d}^{\dagger}}\Big) -\frac{24}{5} g_{1}^{2} \mbox{Tr}\Big({y_{e}  y_{e}^{\dagger}  y_{e}  y_{e}^{\dagger}}\Big) \nonumber \\
 &- \lambda_H \mbox{Tr}\Big({y_{e}  y_{e}^{\dagger}  y_{e}  y_{e}^{\dagger}}\Big) +2 \lambda_H \mbox{Tr}\Big({y_{e}  y_{L}^{\dagger}  y_{L}  y_{e}^{\dagger}}\Big)  - \lambda_H \mbox{Tr}\Big({y_{L}  y_{L}^{\dagger}  y_{L}  y_{L}^{\dagger}}\Big) -\frac{3}{4} \lambda_H \mbox{Tr}\Big({y_{L}  y_{L}^{\dagger}  y_{S}  y_{S}^{\dagger}}\Big) \nonumber \\
 & -\frac{16}{5} g_{1}^{2} \mbox{Tr}\Big({y_{u}  y_{u}^{\dagger}  y_{u}  y_{u}^{\dagger}}\Big) -64 g_{3}^{2} \mbox{Tr}\Big({y_{u}  y_{u}^{\dagger}  y_{u}  y_{u}^{\dagger}}\Big) -3 \lambda_H \mbox{Tr}\Big({y_{u}  y_{u}^{\dagger}  y_{u}  y_{u}^{\dagger}}\Big) \nonumber \\
 & +60 \mbox{Tr}\Big({y_d  y_{d}^{\dagger}  y_d  y_{d}^{\dagger}  y_d  y_{d}^{\dagger}}\Big) -24 \mbox{Tr}\Big({y_d  y_{d}^{\dagger}  y_d  y_{u}^{\dagger}  y_{u}  y_{d}^{\dagger}}\Big) +12 \mbox{Tr}\Big({y_d  y_{u}^{\dagger}  y_{u}  y_{d}^{\dagger}  y_d  y_{d}^{\dagger}}\Big) \nonumber \\
 &-12 \mbox{Tr}\Big({y_d  y_{u}^{\dagger}  y_{u}  y_{u}^{\dagger}  y_{u}  y_{d}^{\dagger}}\Big) +20 \mbox{Tr}\Big({y_{e}  y_{e}^{\dagger}  y_{e}  y_{e}^{\dagger}  y_{e}  y_{e}^{\dagger}}\Big) +8 \mbox{Tr}\Big({y_{e}  y_{e}^{\dagger}  y_{e}  y_{L}^{\dagger}  y_{L}  y_{e}^{\dagger}}\Big) \nonumber \\
 &+20 \mbox{Tr}\Big({y_{e}  y_{L}^{\dagger}  y_{L}  y_{e}^{\dagger}  y_{e}  y_{e}^{\dagger}}\Big) +28 \mbox{Tr}\Big({y_{e}  y_{L}^{\dagger}  y_{L}  y_{L}^{\dagger}  y_{L}  y_{e}^{\dagger}}\Big) +20 \mbox{Tr}\Big({y_{L}  y_{L}^{\dagger}  y_{L}  y_{L}^{\dagger}  y_{L}  y_{L}^{\dagger}}\Big) \nonumber \\
 &+\mbox{Tr}\Big({y_{L}  y_{L}^{\dagger}  y_{L}  y_{L}^{\dagger}  y_{S}  y_{S}^{\dagger}}\Big) +60 \mbox{Tr}\Big({y_{u}  y_{u}^{\dagger}  y_{u}  y_{u}^{\dagger}  y_{u}  y_{u}^{\dagger}}\Big)
\end{align}

\begin{align}
\beta_{\lambda_{HS}}^{(1)} & =
+2 \lambda_{5} \lambda_{HX} -\frac{9}{10} g_{1}^{2} \lambda_{HS} -\frac{9}{2} g_{2}^{2} \lambda_{HS} +6 \lambda_H \lambda_{HS} +4 \lambda_{S} \lambda_{HS} +4 \lambda_{HS}^{2} \nonumber \\
 &+6 \lambda_{HS} \mbox{Tr}\Big({y_d  y_{d}^{\dagger}}\Big) +2 \lambda_{HS} \mbox{Tr}\Big({y_{e}  y_{e}^{\dagger}}\Big) +2 \lambda_{HS} \mbox{Tr}\Big({y_{L}  y_{L}^{\dagger}}\Big) +\frac{1}{2} \lambda_{HS} \mbox{Tr}\Big({y_{S}  y_{S}^{\dagger}}\Big) \nonumber \\
 &+6 \lambda_{HS} \mbox{Tr}\Big({y_{u}  y_{u}^{\dagger}}\Big) - \mbox{Tr}\Big({y_{L}  y_{L}^{\dagger}  y_{S}  y_{S}^{\dagger}}\Big) \\
\beta_{\lambda_{HS}}^{(2)} & =
 -4 \lambda_{5}^{2} \lambda_{HX} -4 \lambda_{5} \lambda_{HX}^{2} +\frac{1671}{400} g_{1}^{4} \lambda_{HS} +\frac{9}{8} g_{1}^{2} g_{2}^{2} \lambda_{HS} -\frac{145}{16} g_{2}^{4} \lambda_{HS} \nonumber \\
 &- \lambda_{5}^{2} \lambda_{HS} +\frac{36}{5} g_{1}^{2} \lambda_H \lambda_{HS} +36 g_{2}^{2} \lambda_H \lambda_{HS} -15 \lambda_H^{2} \lambda_{HS} -8 \lambda_{5} \lambda_{HX} \lambda_{HS} \nonumber \\
 &- \lambda_{HX}^{2} \lambda_{HS} +\frac{3}{5} g_{1}^{2} \lambda_{HS}^{2} +3 g_{2}^{2} \lambda_{HS}^{2} -36 \lambda_H \lambda_{HS}^{2} -24 \lambda_{S} \lambda_{HS}^{2} -11 \lambda_{HS}^{3} \nonumber \\
 &-\frac{9}{2} \Big(2 \lambda_{HX}  + 3 \lambda_{HS} \Big)|\lambda_{J}|^2 +\frac{1}{4} \Big(-144 \lambda_H  + 160 g_{3}^{2}  + 45 g_{2}^{2}  -48 \lambda_{HS}  \nonumber \\
 & +5 g_{1}^{2} \Big)\lambda_{HS} \mbox{Tr}\Big({y_d  y_{d}^{\dagger}}\Big) +\frac{15}{4} g_{1}^{2} \lambda_{HS} \mbox{Tr}\Big({y_{e}  y_{e}^{\dagger}}\Big) +\frac{15}{4} g_{2}^{2} \lambda_{HS} \mbox{Tr}\Big({y_{e}  y_{e}^{\dagger}}\Big) \nonumber \\
 & -12 \lambda_H \lambda_{HS} \mbox{Tr}\Big({y_{e}  y_{e}^{\dagger}}\Big) -4 \lambda_{HS}^{2} \mbox{Tr}\Big({y_{e}  y_{e}^{\dagger}}\Big) \nonumber \\
 &+\frac{3}{4} g_{1}^{2} \lambda_{HS} \mbox{Tr}\Big({y_{L}  y_{L}^{\dagger}}\Big) +\frac{15}{4} g_{2}^{2} \lambda_{HS} \mbox{Tr}\Big({y_{L}  y_{L}^{\dagger}}\Big) -12 \lambda_H \lambda_{HS} \mbox{Tr}\Big({y_{L}  y_{L}^{\dagger}}\Big) -4 \lambda_{HS}^{2} \mbox{Tr}\Big({y_{L}  y_{L}^{\dagger}}\Big) \nonumber \\
 &- 2 \lambda_{S} \lambda_{HS} \mbox{Tr}\Big({y_{S}  y_{S}^{\dagger}}\Big) - \lambda_{HS}^{2} \mbox{Tr}\Big({y_{S}  y_{S}^{\dagger}}\Big) +\frac{17}{4} g_{1}^{2} \lambda_{HS} \mbox{Tr}\Big({y_{u}  y_{u}^{\dagger}}\Big) \nonumber \\
 & +\frac{45}{4} g_{2}^{2} \lambda_{HS} \mbox{Tr}\Big({y_{u}  y_{u}^{\dagger}}\Big) +40 g_{3}^{2} \lambda_{HS} \mbox{Tr}\Big({y_{u}  y_{u}^{\dagger}}\Big) -36 \lambda_H \lambda_{HS} \mbox{Tr}\Big({y_{u}  y_{u}^{\dagger}}\Big) -12 \lambda_{HS}^{2} \mbox{Tr}\Big({y_{u}  y_{u}^{\dagger}}\Big) \nonumber \\
 &-2 \lambda_{5} \lambda_{HX} \mbox{Tr}\Big({y_{X}  y_{X}^*}\Big) -\frac{27}{2} \lambda_{HS} \mbox{Tr}\Big({y_d  y_{d}^{\dagger}  y_d  y_{d}^{\dagger}}\Big) -21 \lambda_{HS} \mbox{Tr}\Big({y_d  y_{u}^{\dagger}  y_{u}  y_{d}^{\dagger}}\Big) \nonumber \\
 &-\frac{9}{2} \lambda_{HS} \mbox{Tr}\Big({y_{e}  y_{e}^{\dagger}  y_{e}  y_{e}^{\dagger}}\Big) +\lambda_{HS} \mbox{Tr}\Big({y_{e}  y_{L}^{\dagger}  y_{L}  y_{e}^{\dagger}}\Big) -\frac{9}{2} \lambda_{HS} \mbox{Tr}\Big({y_{L}  y_{L}^{\dagger}  y_{L}  y_{L}^{\dagger}}\Big) \nonumber \\
 &+\frac{7}{8} \lambda_{HS} \mbox{Tr}\Big({y_{L}  y_{L}^{\dagger}  y_{S}  y_{S}^{\dagger}}\Big) -\frac{3}{16} \lambda_{HS} \mbox{Tr}\Big({y_{S}  y_{S}^{\dagger}  y_{S}  y_{S}^{\dagger}}\Big) \nonumber \\
 &-\frac{3}{8} \lambda_{HS} \mbox{Tr}\Big({y_{S}  y_{X}^*  y_{X}  y_{S}^{\dagger}}\Big) -\frac{27}{2} \lambda_{HS} \mbox{Tr}\Big({y_{u}  y_{u}^{\dagger}  y_{u}  y_{u}^{\dagger}}\Big) \nonumber \\
 &+\frac{7}{2} \mbox{Tr}\Big({y_{e}  y_{L}^{\dagger}  y_{S}  y_{S}^{\dagger}  y_{L}  y_{e}^{\dagger}}\Big) +\frac{5}{2} \mbox{Tr}\Big({y_{L}  y_{L}^{\dagger}  y_{L}  y_{L}^{\dagger}  y_{S}  y_{S}^{\dagger}}\Big) +\mbox{Tr}\Big({y_{L}  y_{L}^{\dagger}  y_{S}  y_{S}^{\dagger}  y_{L}  y_{L}^{\dagger}}\Big) \nonumber \\
 &+\frac{5}{8} \mbox{Tr}\Big({y_{L}  y_{L}^{\dagger}  y_{S}  y_{S}^{\dagger}  y_{S}  y_{S}^{\dagger}}\Big) +\frac{1}{2} \mbox{Tr}\Big({y_{L}  y_{L}^{\dagger}  y_{S}  y_{X}^*  y_{X}  y_{S}^{\dagger}}\Big)
\end{align}

\begin{align}
\beta_{\lambda_{HX}}^{(1)} & =
-\frac{9}{10} g_{1}^{2} \lambda_{HX} -\frac{9}{2} g_{2}^{2} \lambda_{HX} +6 \lambda_H \lambda_{HX} +4 \lambda_{X} \lambda_{HX} +4 \lambda_{HX}^{2} +2 \lambda_{5} \lambda_{HS} \nonumber \\
 &+6 \lambda_{HX} \mbox{Tr}\Big({y_d  y_{d}^{\dagger}}\Big) +2 \lambda_{HX} \mbox{Tr}\Big({y_{e}  y_{e}^{\dagger}}\Big) +2 \lambda_{HX} \mbox{Tr}\Big({y_{L}  y_{L}^{\dagger}}\Big) +6 \lambda_{HX} \mbox{Tr}\Big({y_{u}  y_{u}^{\dagger}}\Big) \nonumber \\
 &+\lambda_{HX} \mbox{Tr}\Big({y_{X}  y_{X}^*}\Big)\\
\beta_{\lambda_{HX}}^{(2)} & =
 +\frac{1671}{400} g_{1}^{4} \lambda_{HX} +\frac{9}{8} g_{1}^{2} g_{2}^{2} \lambda_{HX} -\frac{145}{16} g_{2}^{4} \lambda_{HX} - \lambda_{5}^{2} \lambda_{HX} +\frac{36}{5} g_{1}^{2} \lambda_H \lambda_{HX}  \nonumber \\
 &+ 36 g_{2}^{2} \lambda_H \lambda_{HX} -15 \lambda_H^{2} \lambda_{HX} -10 \lambda_{X}^2 \lambda_{HX} +\frac{3}{5} g_{1}^{2} \lambda_{HX}^{2} +3 g_{2}^{2} \lambda_{HX}^{2} \nonumber \\
 &-36 \lambda_H \lambda_{HX}^{2} -24 \lambda_{X} \lambda_{HX}^{2} -11 \lambda_{HX}^{3} -4 \lambda_{5}^{2} \lambda_{HS} -8 \lambda_{5} \lambda_{HX} \lambda_{HS} \nonumber \\
 &-4 \lambda_{5} \lambda_{HS}^{2} - \lambda_{HX} \lambda_{HS}^{2} +\frac{3}{2} \Big(-6 \lambda_{HS} + \lambda_{HX}\Big)|\lambda_{J}|^2 \nonumber \\
 &+\frac{1}{4} \Big(-144 \lambda_H  + 160 g_{3}^{2}  + 45 g_{2}^{2}  -48 \lambda_{HX} + 5 g_{1}^{2} \Big)\lambda_{HX} \mbox{Tr}\Big({y_d  y_{d}^{\dagger}}\Big) \nonumber \\
 &+ \frac{15}{4} g_{1}^{2} \lambda_{HX} \mbox{Tr}\Big({y_{e}  y_{e}^{\dagger}}\Big) +\frac{15}{4} g_{2}^{2} \lambda_{HX} \mbox{Tr}\Big({y_{e}  y_{e}^{\dagger}}\Big) -12 \lambda_H \lambda_{HX} \mbox{Tr}\Big({y_{e}  y_{e}^{\dagger}}\Big) -4 \lambda_{HX}^{2} \mbox{Tr}\Big({y_{e}  y_{e}^{\dagger}}\Big) \nonumber \\
 &+\frac{3}{4} g_{1}^{2} \lambda_{HX} \mbox{Tr}\Big({y_{L}  y_{L}^{\dagger}}\Big) +\frac{15}{4} g_{2}^{2} \lambda_{HX} \mbox{Tr}\Big({y_{L}  y_{L}^{\dagger}}\Big) -12 \lambda_H \lambda_{HX} \mbox{Tr}\Big({y_{L}  y_{L}^{\dagger}}\Big) -4 \lambda_{HX}^{2} \mbox{Tr}\Big({y_{L}  y_{L}^{\dagger}}\Big) \nonumber\\
 &- \lambda_{5} \lambda_{HS} \mbox{Tr}\Big({y_{S}  y_{S}^{\dagger}}\Big) +\frac{17}{4} g_{1}^{2} \lambda_{HX} \mbox{Tr}\Big({y_{u}  y_{u}^{\dagger}}\Big) +\frac{45}{4} g_{2}^{2} \lambda_{HX} \mbox{Tr}\Big({y_{u}  y_{u}^{\dagger}}\Big) \nonumber \\
 & +40 g_{3}^{2} \lambda_{HX} \mbox{Tr}\Big({y_{u}  y_{u}^{\dagger}}\Big) -36 \lambda_H \lambda_{HX} \mbox{Tr}\Big({y_{u}  y_{u}^{\dagger}}\Big)  -12 \lambda_{HX}^{2} \mbox{Tr}\Big({y_{u}  y_{u}^{\dagger}}\Big) \nonumber \\
 &-4 \lambda_{X} \lambda_{HX} \mbox{Tr}\Big({y_{X}  y_{X}^*}\Big) -2 \lambda_{HX}^{2} \mbox{Tr}\Big({y_{X}  y_{X}^*}\Big) -\frac{27}{2} \lambda_{HX} \mbox{Tr}\Big({y_d  y_{d}^{\dagger}  y_d  y_{d}^{\dagger}}\Big) \nonumber \\
 &-21 \lambda_{HX} \mbox{Tr}\Big({y_d  y_{u}^{\dagger}  y_{u}  y_{d}^{\dagger}}\Big) -\frac{9}{2} \lambda_{HX} \mbox{Tr}\Big({y_{e}  y_{e}^{\dagger}  y_{e}  y_{e}^{\dagger}}\Big) +\lambda_{HX} \mbox{Tr}\Big({y_{e}  y_{L}^{\dagger}  y_{L}  y_{e}^{\dagger}}\Big) \nonumber \\
 &-\frac{9}{2} \lambda_{HX} \mbox{Tr}\Big({y_{L}  y_{L}^{\dagger}  y_{L}  y_{L}^{\dagger}}\Big) -\frac{3}{8} \lambda_{HX} \mbox{Tr}\Big({y_{L}  y_{L}^{\dagger}  y_{S}  y_{S}^{\dagger}}\Big) -\frac{3}{8} \lambda_{HX} \mbox{Tr}\Big({y_{S}  y_{X}^*  y_{X}  y_{S}^{\dagger}}\Big) \nonumber \\
 &-\frac{27}{2} \lambda_{HX} \mbox{Tr}\Big({y_{u}  y_{u}^{\dagger}  y_{u}  y_{u}^{\dagger}}\Big) -\frac{3}{2} \lambda_{HX} \mbox{Tr}\Big({y_{X}  y_{X}^*  y_{X}  y_{X}^*}\Big) +\mbox{Tr}\Big({y_{L}  y_{L}^{\dagger}  y_{S}  y_{X}^*  y_{X}  y_{S}^{\dagger}}\Big)
\end{align} 

\begin{align}
\beta_{\lambda_{S}}^{(1)} & =
10 \lambda_{S}^2  + 2 \lambda_{5}^{2}  + 4 \lambda_{HS}^{2}  + 9 |\lambda_{J}|^2  -\frac{1}{4} \mbox{Tr}\Big({y_{S}  y_{S}^{\dagger}  y_{S}  y_{S}^{\dagger}}\Big)  + \lambda_{S} \mbox{Tr}\Big({y_{S}  y_{S}^{\dagger}}\Big) \\
\beta_{\lambda_{S}}^{(2)} & =
  -8 \lambda_{5}^{3} -10 \lambda_{5}^{2} \lambda_{S} -60 \lambda_H^{\Big(S\Big),3} +\frac{24}{5} g_{1}^{2} \lambda_{HS}^{2} +24 g_{2}^{2} \lambda_{HS}^{2} -20 \lambda_{S} \lambda_{HS}^{2} -16 \lambda_{HS}^{3} \nonumber \\
 &-24 \lambda_{HS}^{2} \mbox{Tr}\Big({y_d  y_{d}^{\dagger}}\Big) -8 \lambda_{HS}^{2} \mbox{Tr}\Big({y_{e}  y_{e}^{\dagger}}\Big)-8 \lambda_{HS}^{2} \mbox{Tr}\Big({y_{L}  y_{L}^{\dagger}}\Big) \nonumber \\
 & -5 \lambda_{S}^2 \mbox{Tr}\Big({y_{S}  y_{S}^{\dagger}}\Big) -24 \lambda_{HS}^{2} \mbox{Tr}\Big({y_{u}  y_{u}^{\dagger}}\Big) +\frac{9}{100} |\lambda_{J}|^2 \Big(-1100 \lambda_{S}  -25 \mbox{Tr}\Big({y_{S}  y_{S}^{\dagger}}\Big) \nonumber \\
 &-50 \mbox{Tr}\Big({y_{X}  y_{X}^*}\Big)  -600 \lambda_{5} \Big)-2 \lambda_{5}^{2} \mbox{Tr}\Big({y_{X}  y_{X}^*}\Big) -\frac{3}{2} \lambda_{S} \mbox{Tr}\Big({y_{L}  y_{L}^{\dagger}  y_{S}  y_{S}^{\dagger}}\Big) \nonumber \\
 &+\frac{1}{8} \lambda_{S} \mbox{Tr}\Big({y_{S}  y_{S}^{\dagger}  y_{S}  y_{S}^{\dagger}}\Big) -\frac{3}{4} \lambda_{S} \mbox{Tr}\Big({y_{S}  y_{X}^*  y_{X}  y_{S}^{\dagger}}\Big) \nonumber \\
 &+\frac{1}{2} \mbox{Tr}\Big({y_{L}  y_{L}^{\dagger}  y_{S}  y_{S}^{\dagger}  y_{S}  y_{S}^{\dagger}}\Big) +\frac{1}{4} \mbox{Tr}\Big({y_{S}  y_{S}^{\dagger}  y_{S}  y_{S}^{\dagger}  y_{S}  y_{S}^{\dagger}}\Big) \nonumber \\
 & +\frac{1}{8} \mbox{Tr}\Big({y_{S}  y_{S}^{\dagger}  y_{S}  y_{X}^*  y_{X}  y_{S}^{\dagger}}\Big) +\frac{1}{8} \mbox{Tr}\Big({y_{S}  y_{X}^*  y_{X}  y_{S}^{\dagger}  y_{S}  y_{S}^{\dagger}}\Big) \nonumber \\
 &+\frac{1}{4} \mbox{Tr}\Big({y_{S}  y_{X}^*  y_{S}^{T}  y_{S}^*  y_{X}  y_{S}^{\dagger}}\Big)
\end{align}

\begin{align}
\beta_{\lambda_{5}}^{(1)} = &
+4 \lambda_{5}^{2} +4 \lambda_{5} \lambda_{S} +4 \lambda_{5} \lambda_{X} +4 \lambda_{HX} \lambda_{HS} +9 |\lambda_{J}|^2 +\frac{1}{2} \lambda_{5} \mbox{Tr}\Big({y_{S}  y_{S}^{\dagger}}\Big)
 +\lambda_{5} \mbox{Tr}\Big({y_{X}  y_{X}^*}\Big) - \mbox{Tr}\Big({y_{S}  y_{X}^*  y_{X}  y_{S}^{\dagger}}\Big) \\
\beta_{\lambda_{5}}^{(2)} = &
 -10 \lambda_{5}^{3} -24 \lambda_{5}^{2} \lambda_{S} -10 \lambda_{5} \lambda_{S}^2 -24 \lambda_{5}^{2} \lambda_{X} -10 \lambda_{5} \lambda_{X}^2 -2 \lambda_{5} \lambda_{HX}^{2} +\frac{24}{5} g_{1}^{2} \lambda_{HX} \lambda_{HS} \nonumber \\
 &  +24 g_{2}^{2} \lambda_{HX} \lambda_{HS} -16 \lambda_{5} \lambda_{HX} \lambda_{HS} -8 \lambda_{HX}^{2} \lambda_{HS} -2 \lambda_{5} \lambda_{HS}^{2} -8 \lambda_{HX} \lambda_{HS}^{2} -24 \lambda_{HX} \lambda_{HS} \mbox{Tr}\Big({y_d  y_{d}^{\dagger}}\Big) \nonumber \\
 &-8 \lambda_{HX} \lambda_{HS} \mbox{Tr}\Big({y_{e}  y_{e}^{\dagger}}\Big) -8 \lambda_{HX} \lambda_{HS} \mbox{Tr}\Big({y_{L}  y_{L}^{\dagger}}\Big) - \lambda_{5}^{2} \mbox{Tr}\Big({y_{S}  y_{S}^{\dagger}}\Big) -2 \lambda_{5} \lambda_{S} \mbox{Tr}\Big({y_{S}  y_{S}^{\dagger}}\Big) \nonumber \\
 &-\frac{3}{50} |\lambda_{J}|^2 \Big(4 \Big(225 \lambda_{S}  + 275 \lambda_{5}  + 75 \lambda_{X} \Big) + 75 \mbox{Tr}\Big({y_{S}  y_{S}^{\dagger}}\Big) \Big)-24 \lambda_{HX} \lambda_{HS} \mbox{Tr}\Big({y_{u}  y_{u}^{\dagger}}\Big)  \nonumber \\
 &-2 \lambda_{5}^{2} \mbox{Tr}\Big({y_{X}  y_{X}^*}\Big) -4 \lambda_{5} \lambda_{X} \mbox{Tr}\Big({y_{X}  y_{X}^*}\Big) -\frac{3}{4} \lambda_{5} \mbox{Tr}\Big({y_{L}  y_{L}^{\dagger}  y_{S}  y_{S}^{\dagger}}\Big) \nonumber \\
 &-\frac{3}{16} \lambda_{5} \mbox{Tr}\Big({y_{S}  y_{S}^{\dagger}  y_{S}  y_{S}^{\dagger}}\Big) +\frac{5}{4} \lambda_{5} \mbox{Tr}\Big({y_{S}  y_{X}^*  y_{X}  y_{S}^{\dagger}}\Big) -\frac{3}{2} \lambda_{5} \mbox{Tr}\Big({y_{X}  y_{X}^*  y_{X}  y_{X}^*}\Big) \nonumber \\
 &+\mbox{Tr}\Big({y_{L}  y_{L}^{\dagger}  y_{S}  y_{X}^*  y_{X}  y_{S}^{\dagger}}\Big)+\frac{9}{32} \mbox{Tr}\Big({y_{S}  y_{S}^{\dagger}  y_{S}  y_{X}^*  y_{X}  y_{S}^{\dagger}}\Big) \nonumber\\
 &+\frac{11}{32} \mbox{Tr}\Big({y_{S}  y_{X}^*  y_{X}  y_{S}^{\dagger}  y_{S}  y_{S}^{\dagger}}\Big) +\frac{5}{2} \mbox{Tr}\Big({y_{S}  y_{X}^*  y_{X}  y_{X}^*  y_{X}  y_{S}^{\dagger}}\Big) +\frac{3}{8} \mbox{Tr}\Big({y_{S}  y_{X}^*  y_{S}^{T}  y_{S}^*  y_{X}  y_{S}^{\dagger}}\Big) 
 \end{align}
 
\begin{align}
\beta_{\lambda_{X}}^{(1)} = & ~ 10 \lambda_{X}^2  + 2 \lambda_{X} \mbox{Tr}\Big({y_{X}  y_{X}^*}\Big)  + 2 \lambda_{5}^{2}  -2 \mbox{Tr}\Big({y_{X}  y_{X}^*  y_{X}  y_{X}^*}\Big)  + 4 \lambda_{HX}^{2} \\
\beta_{\lambda_{X}}^{(2)} = &
 -8 \lambda_{5}^{3} -10 \lambda_{5}^{2} \lambda_{X} -60 \lambda_{X}^3 +\frac{24}{5} g_{1}^{2} \lambda_{HX}^{2} +24 g_{2}^{2} \lambda_{HX}^{2} -20 \lambda_{X} \lambda_{HX}^{2} -16 \lambda_{HX}^{3} +3 \Big(-6 \lambda_{5} + \lambda_{X}\Big)|\lambda_{J}|^2  \nonumber \\
 &  -24 \lambda_{HX}^{2} \mbox{Tr}\Big({y_d  y_{d}^{\dagger}}\Big) -8 \lambda_{HX}^{2} \mbox{Tr}\Big({y_{e}  y_{e}^{\dagger}}\Big)  -8 \lambda_{HX}^{2} \mbox{Tr}\Big({y_{L}  y_{L}^{\dagger}}\Big) - \lambda_{5}^{2} \mbox{Tr}\Big({y_{S}  y_{S}^{\dagger}}\Big) -24 \lambda_{HX}^{2} \mbox{Tr}\Big({y_{u}  y_{u}^{\dagger}}\Big) \nonumber \\
 & -10 \lambda_{X}^2 \mbox{Tr}\Big({y_{X}  y_{X}^*}\Big) -\frac{3}{4} \lambda_{X} \mbox{Tr}\Big({y_{S}  y_{X}^*  y_{X}  y_{S}^{\dagger}}\Big) +\lambda_{X} \mbox{Tr}\Big({y_{X}  y_{X}^*  y_{X}  y_{X}^*}\Big) \nonumber \\
 &  +\mbox{Tr}\Big({y_{S}  y_{X}^*  y_{X}  y_{X}^*  y_{X}  y_{S}^{\dagger}}\Big)+8 \mbox{Tr}\Big({y_{X}  y_{X}^*  y_{X}  y_{X}^*  y_{X}  y_{X}^*}\Big)
\end{align} 

\begin{align}
\beta_{\lambda_{J}}^{(1)} = & ~ 6 \lambda_{S} \lambda_{J}  + 6 \lambda_{5} \lambda_{J} + \frac{1}{2} \lambda_{J} \mbox{Tr}\Big({y_{X}  y_{X}^*}\Big) + \frac{3}{4} \lambda_{J} \mbox{Tr}\Big({y_{S}  y_{S}^{\dagger}}\Big) \\
\beta_{\lambda_{J}}^{(2)}  = & 
-22 \lambda_{5}^{2} \lambda_{J} -36 \lambda_{5} \lambda_{S} \lambda_{J} -33 \lambda_{S}^2 \lambda_{J} -12 \lambda_{5} \lambda_{X} \lambda_{J} +\lambda_{X}^2 \lambda_{J} +\lambda_{J} \lambda_{HX}^{2} -12 \lambda_{J} \lambda_{HX} \lambda_{HS} \nonumber \\
 & -9 \lambda_{J} \lambda_{HS}^{2} +\frac{15}{2} \lambda_{J}^{2} \lambda_{J}^* +\frac{3}{40} \Big(-20 \Big(2 \lambda_{S} + \lambda_{5}\Big) \lambda_{J} \mbox{Tr}\Big({y_{S}  y_{S}^{\dagger}}\Big) -3 \lambda_{5} \lambda_{J} \mbox{Tr}\Big({y_{X}  y_{X}^*}\Big) \nonumber \\
 &-\frac{9}{8} \lambda_{J} \mbox{Tr}\Big({y_{L}  y_{L}^{\dagger}  y_{S}  y_{S}^{\dagger}}\Big) +\frac{27}{32} \lambda_{J} \mbox{Tr}\Big({y_{S}  y_{S}^{\dagger}  y_{S}  y_{S}^{\dagger}}\Big) +3 \lambda_{J} \mbox{Tr}\Big({y_{S}  y_{X}^*  y_{X}  y_{S}^{\dagger}}\Big) -\frac{3}{4} \lambda_{J} \mbox{Tr}\Big({y_{X}  y_{X}^*  y_{X}  y_{X}^*}\Big)
\end{align}

\begin{acknowledgments}
This work was supported by the Spanish MINECO under grants FPA2014-58183-P, and MULTIDARK CSD2009-00064 (Consolider-Ingenio 2010 Programme); by Generalitat Valenciana grant PROMETEOII/2014/084, and Centro de Excelencia Severo Ochoa SEV-2014-0398.
N.~R. was partly funded by becas de postdoctorado en el extranjero Conicyt/Becas Chile 74150028 and by proyecto Fondecyt Postdoctorado (2017) num. 3170135.  \href{http://goo.gl/HFVG6B}{N.~R.} also wants to thank P. Sánchez for her constant support and all the members at the AHEP group for the hospitality.
R.~A.~L. acknowledges the support of the Juan de la Cierva contract JCI-2012-12901 (MINECO), the Spanish MESS via the {\it Servicio Público de Empleo Estatal}, and the UCN's Publication Incentive program No. CPIP20180343.\\
F.G.C. acknowledges the financial support from CONACYT, CONACYT-SNI and PAPIIT IN111115. \\
\end{acknowledgments}

\bibliography{refpaper}

\begin{thebibliography}{67}%
\makeatletter
\providecommand \@ifxundefined [1]{%
 \@ifx{#1\undefined}
}%
\providecommand \@ifnum [1]{%
 \ifnum #1\expandafter \@firstoftwo
 \else \expandafter \@secondoftwo
 \fi
}%
\providecommand \@ifx [1]{%
 \ifx #1\expandafter \@firstoftwo
 \else \expandafter \@secondoftwo
 \fi
}%
\providecommand \natexlab [1]{#1}%
\providecommand \enquote  [1]{``#1''}%
\providecommand \bibnamefont  [1]{#1}%
\providecommand \bibfnamefont [1]{#1}%
\providecommand \citenamefont [1]{#1}%
\providecommand \href@noop [0]{\@secondoftwo}%
\providecommand \href [0]{\begingroup \@sanitize@url \@href}%
\providecommand \@href[1]{\@@startlink{#1}\@@href}%
\providecommand \@@href[1]{\endgroup#1\@@endlink}%
\providecommand \@sanitize@url [0]{\catcode `\\12\catcode `\$12\catcode
  `\&12\catcode `\#12\catcode `\^12\catcode `\_12\catcode `\%12\relax}%
\providecommand \@@startlink[1]{}%
\providecommand \@@endlink[0]{}%
\providecommand \url  [0]{\begingroup\@sanitize@url \@url }%
\providecommand \@url [1]{\endgroup\@href {#1}{\urlprefix }}%
\providecommand \urlprefix  [0]{URL }%
\providecommand \Eprint [0]{\href }%
\providecommand \doibase [0]{http://dx.doi.org/}%
\providecommand \selectlanguage [0]{\@gobble}%
\providecommand \bibinfo  [0]{\@secondoftwo}%
\providecommand \bibfield  [0]{\@secondoftwo}%
\providecommand \translation [1]{[#1]}%
\providecommand \BibitemOpen [0]{}%
\providecommand \bibitemStop [0]{}%
\providecommand \bibitemNoStop [0]{.\EOS\space}%
\providecommand \EOS [0]{\spacefactor3000\relax}%
\providecommand \BibitemShut  [1]{\csname bibitem#1\endcsname}%
\let\auto@bib@innerbib\@empty
\bibitem [{\citenamefont {Weinberg}(1967)}]{Weinberg:1967tq}%
  \BibitemOpen
  \bibfield  {author} {\bibinfo {author} {\bibfnamefont {S.}~\bibnamefont
  {Weinberg}},\ }\href {\doibase 10.1103/PhysRevLett.19.1264} {\bibfield
  {journal} {\bibinfo  {journal} {Phys. Rev. Lett.}\ }\textbf {\bibinfo
  {volume} {19}},\ \bibinfo {pages} {1264} (\bibinfo {year}
  {1967})}\BibitemShut {NoStop}%
\bibitem [{\citenamefont {Higgs}(1964)}]{Higgs:1964pj}%
  \BibitemOpen
  \bibfield  {author} {\bibinfo {author} {\bibfnamefont {P.~W.}\ \bibnamefont
  {Higgs}},\ }\href {\doibase 10.1103/PhysRevLett.13.508} {\bibfield  {journal}
  {\bibinfo  {journal} {Phys. Rev. Lett.}\ }\textbf {\bibinfo {volume} {13}},\
  \bibinfo {pages} {508} (\bibinfo {year} {1964})}\BibitemShut {NoStop}%
\bibitem [{\citenamefont {Arnison}\ \emph {et~al.}(1983)\citenamefont {Arnison}
  \emph {et~al.}}]{Arnison:1983rp}%
  \BibitemOpen
  \bibfield  {author} {\bibinfo {author} {\bibfnamefont {G.}~\bibnamefont
  {Arnison}} \emph {et~al.} (\bibinfo {collaboration} {UA1}),\ }\bibfield
  {booktitle} {\emph {\bibinfo {booktitle} {{Proceedings, 18th Rencontres de
  Moriond on Gluons and Heavy Flavours: La Plagne, France, January 23-29,
  1983}}},\ }\href {\doibase 10.1016/0370-2693(83)91177-2} {\bibfield
  {journal} {\bibinfo  {journal} {Phys. Lett.}\ }\textbf {\bibinfo {volume}
  {122B}},\ \bibinfo {pages} {103} (\bibinfo {year} {1983})},\ \bibinfo {note}
  {[,611(1983)]}\BibitemShut {NoStop}%
\bibitem [{\citenamefont {Bagnaia}\ \emph {et~al.}(1983)\citenamefont {Bagnaia}
  \emph {et~al.}}]{Bagnaia:1983zx}%
  \BibitemOpen
  \bibfield  {author} {\bibinfo {author} {\bibfnamefont {P.}~\bibnamefont
  {Bagnaia}} \emph {et~al.} (\bibinfo {collaboration} {UA2}),\ }\href {\doibase
  10.1016/0370-2693(83)90744-X} {\bibfield  {journal} {\bibinfo  {journal}
  {Phys. Lett.}\ }\textbf {\bibinfo {volume} {B129}},\ \bibinfo {pages} {130}
  (\bibinfo {year} {1983})}\BibitemShut {NoStop}%
\bibitem [{\citenamefont {Aad}\ \emph {et~al.}(2012)\citenamefont {Aad} \emph
  {et~al.}}]{Aad:2012tfa}%
  \BibitemOpen
  \bibfield  {author} {\bibinfo {author} {\bibfnamefont {G.}~\bibnamefont
  {Aad}} \emph {et~al.} (\bibinfo {collaboration} {ATLAS}),\ }\href {\doibase
  10.1016/j.physletb.2012.08.020} {\bibfield  {journal} {\bibinfo  {journal}
  {Phys. Lett.}\ }\textbf {\bibinfo {volume} {B716}},\ \bibinfo {pages} {1}
  (\bibinfo {year} {2012})},\ \Eprint {http://arxiv.org/abs/1207.7214}
  {arXiv:1207.7214 [hep-ex]} \BibitemShut {NoStop}%
\bibitem [{\citenamefont {Chatrchyan}\ \emph {et~al.}(2012)\citenamefont
  {Chatrchyan} \emph {et~al.}}]{Chatrchyan:2012xdj}%
  \BibitemOpen
  \bibfield  {author} {\bibinfo {author} {\bibfnamefont {S.}~\bibnamefont
  {Chatrchyan}} \emph {et~al.} (\bibinfo {collaboration} {CMS}),\ }\href
  {\doibase 10.1016/j.physletb.2012.08.021} {\bibfield  {journal} {\bibinfo
  {journal} {Phys. Lett.}\ }\textbf {\bibinfo {volume} {B716}},\ \bibinfo
  {pages} {30} (\bibinfo {year} {2012})},\ \Eprint
  {http://arxiv.org/abs/1207.7235} {arXiv:1207.7235 [hep-ex]} \BibitemShut
  {NoStop}%
\bibitem [{\citenamefont {{Oort}}(1932)}]{Oort:1932BAN}%
  \BibitemOpen
  \bibfield  {author} {\bibinfo {author} {\bibfnamefont {J.~H.}\ \bibnamefont
  {{Oort}}},\ }\href@noop {} {\bibfield  {journal} {\bibinfo  {journal} {bain}\
  }\textbf {\bibinfo {volume} {6}},\ \bibinfo {pages} {249} (\bibinfo {year}
  {1932})}\BibitemShut {NoStop}%
\bibitem [{\citenamefont {Rubin}\ \emph {et~al.}(1980)\citenamefont {Rubin},
  \citenamefont {Thonnard},\ and\ \citenamefont {Ford}}]{Rubin:1980zd}%
  \BibitemOpen
  \bibfield  {author} {\bibinfo {author} {\bibfnamefont {V.~C.}\ \bibnamefont
  {Rubin}}, \bibinfo {author} {\bibfnamefont {N.}~\bibnamefont {Thonnard}}, \
  and\ \bibinfo {author} {\bibfnamefont {W.~K.}\ \bibnamefont {Ford},
  \bibfnamefont {Jr.}},\ }\href {\doibase 10.1086/158003} {\bibfield  {journal}
  {\bibinfo  {journal} {Astrophys. J.}\ }\textbf {\bibinfo {volume} {238}},\
  \bibinfo {pages} {471} (\bibinfo {year} {1980})}\BibitemShut {NoStop}%
\bibitem [{\citenamefont {Ade}\ \emph {et~al.}(2016)\citenamefont {Ade} \emph
  {et~al.}}]{Ade:2015xua}%
  \BibitemOpen
  \bibfield  {author} {\bibinfo {author} {\bibfnamefont {P.~A.~R.}\
  \bibnamefont {Ade}} \emph {et~al.} (\bibinfo {collaboration} {Planck}),\
  }\href {\doibase 10.1051/0004-6361/201525830} {\bibfield  {journal} {\bibinfo
   {journal} {Astron. Astrophys.}\ }\textbf {\bibinfo {volume} {594}},\
  \bibinfo {pages} {A13} (\bibinfo {year} {2016})},\ \Eprint
  {http://arxiv.org/abs/1502.01589} {arXiv:1502.01589 [astro-ph.CO]}
  \BibitemShut {NoStop}%
\bibitem [{\citenamefont {Pontecorvo}(1957)}]{Pontecorvo:1957cp}%
  \BibitemOpen
  \bibfield  {author} {\bibinfo {author} {\bibfnamefont {B.}~\bibnamefont
  {Pontecorvo}},\ }\href@noop {} {\bibfield  {journal} {\bibinfo  {journal}
  {Sov. Phys. JETP}\ }\textbf {\bibinfo {volume} {6}},\ \bibinfo {pages} {429}
  (\bibinfo {year} {1957})},\ \bibinfo {note} {[Zh. Eksp. Teor.
  Fiz.33,549(1957)]}\BibitemShut {NoStop}%
\bibitem [{\citenamefont {Maki}\ \emph {et~al.}(1962)\citenamefont {Maki},
  \citenamefont {Nakagawa},\ and\ \citenamefont {Sakata}}]{Maki:1962mu}%
  \BibitemOpen
  \bibfield  {author} {\bibinfo {author} {\bibfnamefont {Z.}~\bibnamefont
  {Maki}}, \bibinfo {author} {\bibfnamefont {M.}~\bibnamefont {Nakagawa}}, \
  and\ \bibinfo {author} {\bibfnamefont {S.}~\bibnamefont {Sakata}},\ }\href
  {\doibase 10.1143/PTP.28.870} {\bibfield  {journal} {\bibinfo  {journal}
  {Prog. Theor. Phys.}\ }\textbf {\bibinfo {volume} {28}},\ \bibinfo {pages}
  {870} (\bibinfo {year} {1962})}\BibitemShut {NoStop}%
\bibitem [{\citenamefont {Ahmad}\ \emph {et~al.}(2001)\citenamefont {Ahmad}
  \emph {et~al.}}]{Ahmad:2001an}%
  \BibitemOpen
  \bibfield  {author} {\bibinfo {author} {\bibfnamefont {Q.~R.}\ \bibnamefont
  {Ahmad}} \emph {et~al.} (\bibinfo {collaboration} {SNO}),\ }\href {\doibase
  10.1103/PhysRevLett.87.071301} {\bibfield  {journal} {\bibinfo  {journal}
  {Phys. Rev. Lett.}\ }\textbf {\bibinfo {volume} {87}},\ \bibinfo {pages}
  {071301} (\bibinfo {year} {2001})},\ \Eprint
  {http://arxiv.org/abs/nucl-ex/0106015} {arXiv:nucl-ex/0106015 [nucl-ex]}
  \BibitemShut {NoStop}%
\bibitem [{\citenamefont {Fukuda}\ \emph {et~al.}(1998)\citenamefont {Fukuda}
  \emph {et~al.}}]{Fukuda:1998tw}%
  \BibitemOpen
  \bibfield  {author} {\bibinfo {author} {\bibfnamefont {Y.}~\bibnamefont
  {Fukuda}} \emph {et~al.} (\bibinfo {collaboration} {Super-Kamiokande}),\
  }\href {\doibase 10.1016/S0370-2693(98)00476-6} {\bibfield  {journal}
  {\bibinfo  {journal} {Phys. Lett.}\ }\textbf {\bibinfo {volume} {B433}},\
  \bibinfo {pages} {9} (\bibinfo {year} {1998})},\ \Eprint
  {http://arxiv.org/abs/hep-ex/9803006} {arXiv:hep-ex/9803006 [hep-ex]}
  \BibitemShut {NoStop}%
\bibitem [{\citenamefont {Patrignani}\ \emph {et~al.}(2016)\citenamefont
  {Patrignani} \emph {et~al.}}]{Olive:2016xmw}%
  \BibitemOpen
  \bibfield  {author} {\bibinfo {author} {\bibfnamefont {C.}~\bibnamefont
  {Patrignani}} \emph {et~al.} (\bibinfo {collaboration} {Particle Data
  Group}),\ }\href {\doibase 10.1088/1674-1137/40/10/100001} {\bibfield
  {journal} {\bibinfo  {journal} {Chin. Phys.}\ }\textbf {\bibinfo {volume}
  {C40}},\ \bibinfo {pages} {100001} (\bibinfo {year} {2016})}\BibitemShut
  {NoStop}%
\bibitem [{\citenamefont {Esteban}\ \emph {et~al.}(2017)\citenamefont
  {Esteban}, \citenamefont {Gonzalez-Garcia}, \citenamefont {Maltoni},
  \citenamefont {Martinez-Soler},\ and\ \citenamefont
  {Schwetz}}]{Esteban:2016qun}%
  \BibitemOpen
  \bibfield  {author} {\bibinfo {author} {\bibfnamefont {I.}~\bibnamefont
  {Esteban}}, \bibinfo {author} {\bibfnamefont {M.~C.}\ \bibnamefont
  {Gonzalez-Garcia}}, \bibinfo {author} {\bibfnamefont {M.}~\bibnamefont
  {Maltoni}}, \bibinfo {author} {\bibfnamefont {I.}~\bibnamefont
  {Martinez-Soler}}, \ and\ \bibinfo {author} {\bibfnamefont {T.}~\bibnamefont
  {Schwetz}},\ }\href {\doibase 10.1007/JHEP01(2017)087} {\bibfield  {journal}
  {\bibinfo  {journal} {JHEP}\ }\textbf {\bibinfo {volume} {01}},\ \bibinfo
  {pages} {087} (\bibinfo {year} {2017})},\ \Eprint
  {http://arxiv.org/abs/1611.01514} {arXiv:1611.01514 [hep-ph]} \BibitemShut
  {NoStop}%
\bibitem [{\citenamefont {Capozzi}\ \emph {et~al.}(2016)\citenamefont
  {Capozzi}, \citenamefont {Lisi}, \citenamefont {Marrone}, \citenamefont
  {Montanino},\ and\ \citenamefont {Palazzo}}]{Capozzi:2016rtj}%
  \BibitemOpen
  \bibfield  {author} {\bibinfo {author} {\bibfnamefont {F.}~\bibnamefont
  {Capozzi}}, \bibinfo {author} {\bibfnamefont {E.}~\bibnamefont {Lisi}},
  \bibinfo {author} {\bibfnamefont {A.}~\bibnamefont {Marrone}}, \bibinfo
  {author} {\bibfnamefont {D.}~\bibnamefont {Montanino}}, \ and\ \bibinfo
  {author} {\bibfnamefont {A.}~\bibnamefont {Palazzo}},\ }\href {\doibase
  10.1016/j.nuclphysb.2016.02.016} {\bibfield  {journal} {\bibinfo  {journal}
  {Nucl. Phys.}\ }\textbf {\bibinfo {volume} {B908}},\ \bibinfo {pages} {218}
  (\bibinfo {year} {2016})},\ \Eprint {http://arxiv.org/abs/1601.07777}
  {arXiv:1601.07777 [hep-ph]} \BibitemShut {NoStop}%
\bibitem [{\citenamefont {Forero}\ \emph {et~al.}(2014)\citenamefont {Forero},
  \citenamefont {Tortola},\ and\ \citenamefont {Valle}}]{Forero:2014bxa}%
  \BibitemOpen
  \bibfield  {author} {\bibinfo {author} {\bibfnamefont {D.~V.}\ \bibnamefont
  {Forero}}, \bibinfo {author} {\bibfnamefont {M.}~\bibnamefont {Tortola}}, \
  and\ \bibinfo {author} {\bibfnamefont {J.~W.~F.}\ \bibnamefont {Valle}},\
  }\href {\doibase 10.1103/PhysRevD.90.093006} {\bibfield  {journal} {\bibinfo
  {journal} {Phys. Rev.}\ }\textbf {\bibinfo {volume} {D90}},\ \bibinfo {pages}
  {093006} (\bibinfo {year} {2014})},\ \Eprint {http://arxiv.org/abs/1405.7540}
  {arXiv:1405.7540 [hep-ph]} \BibitemShut {NoStop}%
\bibitem [{\citenamefont {Lattanzi}\ \emph {et~al.}(2014)\citenamefont
  {Lattanzi}, \citenamefont {Lineros},\ and\ \citenamefont
  {Taoso}}]{Lattanzi:2014mia}%
  \BibitemOpen
  \bibfield  {author} {\bibinfo {author} {\bibfnamefont {M.}~\bibnamefont
  {Lattanzi}}, \bibinfo {author} {\bibfnamefont {R.~A.}\ \bibnamefont
  {Lineros}}, \ and\ \bibinfo {author} {\bibfnamefont {M.}~\bibnamefont
  {Taoso}},\ }\href {\doibase 10.1088/1367-2630/16/12/125012} {\bibfield
  {journal} {\bibinfo  {journal} {New J. Phys.}\ }\textbf {\bibinfo {volume}
  {16}},\ \bibinfo {pages} {125012} (\bibinfo {year} {2014})},\ \Eprint
  {http://arxiv.org/abs/1406.0004} {arXiv:1406.0004 [hep-ph]} \BibitemShut
  {NoStop}%
\bibitem [{\citenamefont {Hirsch}\ \emph {et~al.}(2013)\citenamefont {Hirsch},
  \citenamefont {Lineros}, \citenamefont {Morisi}, \citenamefont {Palacio},
  \citenamefont {Rojas},\ and\ \citenamefont {Valle}}]{Hirsch:2013ola}%
  \BibitemOpen
  \bibfield  {author} {\bibinfo {author} {\bibfnamefont {M.}~\bibnamefont
  {Hirsch}}, \bibinfo {author} {\bibfnamefont {R.~A.}\ \bibnamefont {Lineros}},
  \bibinfo {author} {\bibfnamefont {S.}~\bibnamefont {Morisi}}, \bibinfo
  {author} {\bibfnamefont {J.}~\bibnamefont {Palacio}}, \bibinfo {author}
  {\bibfnamefont {N.}~\bibnamefont {Rojas}}, \ and\ \bibinfo {author}
  {\bibfnamefont {J.~W.~F.}\ \bibnamefont {Valle}},\ }\href {\doibase
  10.1007/JHEP10(2013)149} {\bibfield  {journal} {\bibinfo  {journal} {JHEP}\
  }\textbf {\bibinfo {volume} {10}},\ \bibinfo {pages} {149} (\bibinfo {year}
  {2013})},\ \Eprint {http://arxiv.org/abs/1307.8134} {arXiv:1307.8134
  [hep-ph]} \BibitemShut {NoStop}%
\bibitem [{\citenamefont {Heurtier}\ and\ \citenamefont
  {Teresi}(2016)}]{Heurtier:2016iac}%
  \BibitemOpen
  \bibfield  {author} {\bibinfo {author} {\bibfnamefont {L.}~\bibnamefont
  {Heurtier}}\ and\ \bibinfo {author} {\bibfnamefont {D.}~\bibnamefont
  {Teresi}},\ }\href {\doibase 10.1103/PhysRevD.94.125022} {\bibfield
  {journal} {\bibinfo  {journal} {Phys. Rev.}\ }\textbf {\bibinfo {volume}
  {D94}},\ \bibinfo {pages} {125022} (\bibinfo {year} {2016})},\ \Eprint
  {http://arxiv.org/abs/1607.01798} {arXiv:1607.01798 [hep-ph]} \BibitemShut
  {NoStop}%
\bibitem [{\citenamefont {de~Salas}\ \emph {et~al.}(2016)\citenamefont
  {de~Salas}, \citenamefont {Lineros},\ and\ \citenamefont
  {Tórtola}}]{deSalas:2016svi}%
  \BibitemOpen
  \bibfield  {author} {\bibinfo {author} {\bibfnamefont {P.~F.}\ \bibnamefont
  {de~Salas}}, \bibinfo {author} {\bibfnamefont {R.~A.}\ \bibnamefont
  {Lineros}}, \ and\ \bibinfo {author} {\bibfnamefont {M.}~\bibnamefont
  {Tórtola}},\ }\href {\doibase 10.1103/PhysRevD.94.123001} {\bibfield
  {journal} {\bibinfo  {journal} {Phys. Rev.}\ }\textbf {\bibinfo {volume}
  {D94}},\ \bibinfo {pages} {123001} (\bibinfo {year} {2016})},\ \Eprint
  {http://arxiv.org/abs/1601.05798} {arXiv:1601.05798 [astro-ph.HE]}
  \BibitemShut {NoStop}%
\bibitem [{\citenamefont {Weinberg}(1979)}]{Weinberg:1979sa}%
  \BibitemOpen
  \bibfield  {author} {\bibinfo {author} {\bibfnamefont {S.}~\bibnamefont
  {Weinberg}},\ }\href {\doibase 10.1103/PhysRevLett.43.1566} {\bibfield
  {journal} {\bibinfo  {journal} {Phys. Rev. Lett.}\ }\textbf {\bibinfo
  {volume} {43}},\ \bibinfo {pages} {1566} (\bibinfo {year}
  {1979})}\BibitemShut {NoStop}%
\bibitem [{\citenamefont {Minkowski}(1977)}]{Minkowski:1977sc}%
  \BibitemOpen
  \bibfield  {author} {\bibinfo {author} {\bibfnamefont {P.}~\bibnamefont
  {Minkowski}},\ }\href {\doibase 10.1016/0370-2693(77)90435-X} {\bibfield
  {journal} {\bibinfo  {journal} {Phys. Lett.}\ }\textbf {\bibinfo {volume}
  {B67}},\ \bibinfo {pages} {421} (\bibinfo {year} {1977})}\BibitemShut
  {NoStop}%
\bibitem [{\citenamefont {Mohapatra}\ and\ \citenamefont
  {Senjanovic}(1980)}]{Mohapatra:1979ia}%
  \BibitemOpen
  \bibfield  {author} {\bibinfo {author} {\bibfnamefont {R.~N.}\ \bibnamefont
  {Mohapatra}}\ and\ \bibinfo {author} {\bibfnamefont {G.}~\bibnamefont
  {Senjanovic}},\ }\href {\doibase 10.1103/PhysRevLett.44.912} {\bibfield
  {journal} {\bibinfo  {journal} {Phys. Rev. Lett.}\ }\textbf {\bibinfo
  {volume} {44}},\ \bibinfo {pages} {912} (\bibinfo {year} {1980})}\BibitemShut
  {NoStop}%
\bibitem [{\citenamefont {Schechter}\ and\ \citenamefont
  {Valle}(1980)}]{Schechter:1980gr}%
  \BibitemOpen
  \bibfield  {author} {\bibinfo {author} {\bibfnamefont {J.}~\bibnamefont
  {Schechter}}\ and\ \bibinfo {author} {\bibfnamefont {J.~W.~F.}\ \bibnamefont
  {Valle}},\ }\href {\doibase 10.1103/PhysRevD.22.2227} {\bibfield  {journal}
  {\bibinfo  {journal} {Phys. Rev.}\ }\textbf {\bibinfo {volume} {D22}},\
  \bibinfo {pages} {2227} (\bibinfo {year} {1980})}\BibitemShut {NoStop}%
\bibitem [{\citenamefont {Mohapatra}\ and\ \citenamefont
  {Senjanovic}(1981)}]{Mohapatra:1980yp}%
  \BibitemOpen
  \bibfield  {author} {\bibinfo {author} {\bibfnamefont {R.~N.}\ \bibnamefont
  {Mohapatra}}\ and\ \bibinfo {author} {\bibfnamefont {G.}~\bibnamefont
  {Senjanovic}},\ }\href {\doibase 10.1103/PhysRevD.23.165} {\bibfield
  {journal} {\bibinfo  {journal} {Phys. Rev.}\ }\textbf {\bibinfo {volume}
  {D23}},\ \bibinfo {pages} {165} (\bibinfo {year} {1981})}\BibitemShut
  {NoStop}%
\bibitem [{\citenamefont {Foot}\ \emph {et~al.}(1989)\citenamefont {Foot},
  \citenamefont {Lew}, \citenamefont {He},\ and\ \citenamefont
  {Joshi}}]{Foot:1988aq}%
  \BibitemOpen
  \bibfield  {author} {\bibinfo {author} {\bibfnamefont {R.}~\bibnamefont
  {Foot}}, \bibinfo {author} {\bibfnamefont {H.}~\bibnamefont {Lew}}, \bibinfo
  {author} {\bibfnamefont {X.~G.}\ \bibnamefont {He}}, \ and\ \bibinfo {author}
  {\bibfnamefont {G.~C.}\ \bibnamefont {Joshi}},\ }\href {\doibase
  10.1007/BF01415558} {\bibfield  {journal} {\bibinfo  {journal} {Z. Phys.}\
  }\textbf {\bibinfo {volume} {C44}},\ \bibinfo {pages} {441} (\bibinfo {year}
  {1989})}\BibitemShut {NoStop}%
\bibitem [{\citenamefont {Riazuddin}\ \emph {et~al.}(1981)\citenamefont
  {Riazuddin}, \citenamefont {Marshak},\ and\ \citenamefont
  {Mohapatra}}]{Riazuddin:1981hz}%
  \BibitemOpen
  \bibfield  {author} {\bibinfo {author} {\bibnamefont {Riazuddin}}, \bibinfo
  {author} {\bibfnamefont {R.~E.}\ \bibnamefont {Marshak}}, \ and\ \bibinfo
  {author} {\bibfnamefont {R.~N.}\ \bibnamefont {Mohapatra}},\ }\href {\doibase
  10.1103/PhysRevD.24.1310} {\bibfield  {journal} {\bibinfo  {journal} {Phys.
  Rev.}\ }\textbf {\bibinfo {volume} {D24}},\ \bibinfo {pages} {1310} (\bibinfo
  {year} {1981})}\BibitemShut {NoStop}%
\bibitem [{\citenamefont {Chikashige}\ \emph {et~al.}(1981)\citenamefont
  {Chikashige}, \citenamefont {Mohapatra},\ and\ \citenamefont
  {Peccei}}]{Chikashige:1980ui}%
  \BibitemOpen
  \bibfield  {author} {\bibinfo {author} {\bibfnamefont {Y.}~\bibnamefont
  {Chikashige}}, \bibinfo {author} {\bibfnamefont {R.~N.}\ \bibnamefont
  {Mohapatra}}, \ and\ \bibinfo {author} {\bibfnamefont {R.}~\bibnamefont
  {Peccei}},\ }\href {\doibase 10.1016/0370-2693(81)90011-3} {\bibfield
  {journal} {\bibinfo  {journal} {Phys. Lett. B}\ }\textbf {\bibinfo {volume}
  {98}},\ \bibinfo {pages} {265} (\bibinfo {year} {1981})}\BibitemShut
  {NoStop}%
\bibitem [{\citenamefont {Rothstein}\ \emph {et~al.}(1993)\citenamefont
  {Rothstein}, \citenamefont {Babu},\ and\ \citenamefont
  {Seckel}}]{Rothstein:1992rh}%
  \BibitemOpen
  \bibfield  {author} {\bibinfo {author} {\bibfnamefont {I.~Z.}\ \bibnamefont
  {Rothstein}}, \bibinfo {author} {\bibfnamefont {K.~S.}\ \bibnamefont {Babu}},
  \ and\ \bibinfo {author} {\bibfnamefont {D.}~\bibnamefont {Seckel}},\ }\href
  {\doibase 10.1016/0550-3213(93)90368-Y} {\bibfield  {journal} {\bibinfo
  {journal} {Nucl. Phys.}\ }\textbf {\bibinfo {volume} {B403}},\ \bibinfo
  {pages} {725} (\bibinfo {year} {1993})},\ \Eprint
  {http://arxiv.org/abs/hep-ph/9301213} {arXiv:hep-ph/9301213 [hep-ph]}
  \BibitemShut {NoStop}%
\bibitem [{\citenamefont {Boucenna}\ \emph {et~al.}(2013)\citenamefont
  {Boucenna}, \citenamefont {Lineros},\ and\ \citenamefont
  {Valle}}]{Boucenna:2012rc}%
  \BibitemOpen
  \bibfield  {author} {\bibinfo {author} {\bibfnamefont {M.~S.}\ \bibnamefont
  {Boucenna}}, \bibinfo {author} {\bibfnamefont {R.~A.}\ \bibnamefont
  {Lineros}}, \ and\ \bibinfo {author} {\bibfnamefont {J.~W.~F.}\ \bibnamefont
  {Valle}},\ }\href {\doibase 10.3389/fphy.2013.00034} {\bibfield  {journal}
  {\bibinfo  {journal} {Front.in Phys.}\ }\textbf {\bibinfo {volume} {1}},\
  \bibinfo {pages} {34} (\bibinfo {year} {2013})},\ \Eprint
  {http://arxiv.org/abs/1204.2576} {arXiv:1204.2576 [hep-ph]} \BibitemShut
  {NoStop}%
\bibitem [{\citenamefont {Berezinsky}\ and\ \citenamefont
  {Valle}(1993)}]{Berezinsky:1993fm}%
  \BibitemOpen
  \bibfield  {author} {\bibinfo {author} {\bibfnamefont {V.}~\bibnamefont
  {Berezinsky}}\ and\ \bibinfo {author} {\bibfnamefont {J.~W.~F.}\ \bibnamefont
  {Valle}},\ }\href {\doibase 10.1016/0370-2693(93)90140-D} {\bibfield
  {journal} {\bibinfo  {journal} {Phys. Lett.}\ }\textbf {\bibinfo {volume}
  {B318}},\ \bibinfo {pages} {360} (\bibinfo {year} {1993})},\ \Eprint
  {http://arxiv.org/abs/hep-ph/9309214} {arXiv:hep-ph/9309214 [hep-ph]}
  \BibitemShut {NoStop}%
\bibitem [{\citenamefont {Lattanzi}\ and\ \citenamefont
  {Valle}(2007)}]{Lattanzi:2007ux}%
  \BibitemOpen
  \bibfield  {author} {\bibinfo {author} {\bibfnamefont {M.}~\bibnamefont
  {Lattanzi}}\ and\ \bibinfo {author} {\bibfnamefont {J.~W.~F.}\ \bibnamefont
  {Valle}},\ }\href {\doibase 10.1103/PhysRevLett.99.121301} {\bibfield
  {journal} {\bibinfo  {journal} {Phys. Rev. Lett.}\ }\textbf {\bibinfo
  {volume} {99}},\ \bibinfo {pages} {121301} (\bibinfo {year} {2007})},\
  \Eprint {http://arxiv.org/abs/0705.2406} {arXiv:0705.2406 [astro-ph]}
  \BibitemShut {NoStop}%
\bibitem [{\citenamefont {Aranda}\ and\ \citenamefont
  {de~Anda}(2010)}]{Aranda:2009yb}%
  \BibitemOpen
  \bibfield  {author} {\bibinfo {author} {\bibfnamefont {A.}~\bibnamefont
  {Aranda}}\ and\ \bibinfo {author} {\bibfnamefont {F.~J.}\ \bibnamefont
  {de~Anda}},\ }\href {\doibase 10.1016/j.physletb.2009.12.026} {\bibfield
  {journal} {\bibinfo  {journal} {Phys. Lett.}\ }\textbf {\bibinfo {volume}
  {B683}},\ \bibinfo {pages} {183} (\bibinfo {year} {2010})},\ \Eprint
  {http://arxiv.org/abs/0909.2667} {arXiv:0909.2667 [hep-ph]} \BibitemShut
  {NoStop}%
\bibitem [{\citenamefont {Queiroz}\ and\ \citenamefont
  {Sinha}(2014)}]{Queiroz:2014yna}%
  \BibitemOpen
  \bibfield  {author} {\bibinfo {author} {\bibfnamefont {F.~S.}\ \bibnamefont
  {Queiroz}}\ and\ \bibinfo {author} {\bibfnamefont {K.}~\bibnamefont
  {Sinha}},\ }\href {\doibase 10.1016/j.physletb.2014.06.016} {\bibfield
  {journal} {\bibinfo  {journal} {Phys. Lett.}\ }\textbf {\bibinfo {volume}
  {B735}},\ \bibinfo {pages} {69} (\bibinfo {year} {2014})},\ \Eprint
  {http://arxiv.org/abs/1404.1400} {arXiv:1404.1400 [hep-ph]} \BibitemShut
  {NoStop}%
\bibitem [{\citenamefont {Wang}\ and\ \citenamefont
  {Han}(2016)}]{Wang:2016vfj}%
  \BibitemOpen
  \bibfield  {author} {\bibinfo {author} {\bibfnamefont {W.}~\bibnamefont
  {Wang}}\ and\ \bibinfo {author} {\bibfnamefont {Z.-L.}\ \bibnamefont {Han}},\
  }\href {\doibase 10.1103/PhysRevD.94.053015} {\bibfield  {journal} {\bibinfo
  {journal} {Phys. Rev.}\ }\textbf {\bibinfo {volume} {D94}},\ \bibinfo {pages}
  {053015} (\bibinfo {year} {2016})},\ \Eprint
  {http://arxiv.org/abs/1605.00239} {arXiv:1605.00239 [hep-ph]} \BibitemShut
  {NoStop}%
\bibitem [{\citenamefont {Garcia-Cely}\ and\ \citenamefont
  {Heeck}(2017)}]{Garcia-Cely:2017oco}%
  \BibitemOpen
  \bibfield  {author} {\bibinfo {author} {\bibfnamefont {C.}~\bibnamefont
  {Garcia-Cely}}\ and\ \bibinfo {author} {\bibfnamefont {J.}~\bibnamefont
  {Heeck}},\ }\href {\doibase 10.1007/JHEP05(2017)102} {\bibfield  {journal}
  {\bibinfo  {journal} {JHEP}\ }\textbf {\bibinfo {volume} {05}},\ \bibinfo
  {pages} {102} (\bibinfo {year} {2017})},\ \Eprint
  {http://arxiv.org/abs/1701.07209} {arXiv:1701.07209 [hep-ph]} \BibitemShut
  {NoStop}%
\bibitem [{\citenamefont {Mohapatra}(1986)}]{Mohapatra:1986aw}%
  \BibitemOpen
  \bibfield  {author} {\bibinfo {author} {\bibfnamefont {R.~N.}\ \bibnamefont
  {Mohapatra}},\ }\href {\doibase 10.1103/PhysRevLett.56.561} {\bibfield
  {journal} {\bibinfo  {journal} {Phys. Rev. Lett.}\ }\textbf {\bibinfo
  {volume} {56}},\ \bibinfo {pages} {561} (\bibinfo {year} {1986})}\BibitemShut
  {NoStop}%
\bibitem [{\citenamefont {Mohapatra}\ and\ \citenamefont
  {Valle}(1986)}]{Mohapatra:1986bd}%
  \BibitemOpen
  \bibfield  {author} {\bibinfo {author} {\bibfnamefont {R.~N.}\ \bibnamefont
  {Mohapatra}}\ and\ \bibinfo {author} {\bibfnamefont {J.~W.~F.}\ \bibnamefont
  {Valle}},\ }\bibfield  {booktitle} {\emph {\bibinfo {booktitle}
  {{Proceedings, 23RD International Conference on High Energy Physics, JULY
  16-23, 1986, Berkeley, CA}}},\ }\href {\doibase 10.1103/PhysRevD.34.1642}
  {\bibfield  {journal} {\bibinfo  {journal} {Phys. Rev.}\ }\textbf {\bibinfo
  {volume} {D34}},\ \bibinfo {pages} {1642} (\bibinfo {year}
  {1986})}\BibitemShut {NoStop}%
\bibitem [{\citenamefont {Barr}(2004)}]{Barr:2003nn}%
  \BibitemOpen
  \bibfield  {author} {\bibinfo {author} {\bibfnamefont {S.~M.}\ \bibnamefont
  {Barr}},\ }\href {\doibase 10.1103/PhysRevLett.92.101601} {\bibfield
  {journal} {\bibinfo  {journal} {Phys. Rev. Lett.}\ }\textbf {\bibinfo
  {volume} {92}},\ \bibinfo {pages} {101601} (\bibinfo {year} {2004})},\
  \Eprint {http://arxiv.org/abs/hep-ph/0309152} {arXiv:hep-ph/0309152 [hep-ph]}
  \BibitemShut {NoStop}%
\bibitem [{\citenamefont {Abada}\ and\ \citenamefont
  {Lucente}(2014)}]{Abada:2014vea}%
  \BibitemOpen
  \bibfield  {author} {\bibinfo {author} {\bibfnamefont {A.}~\bibnamefont
  {Abada}}\ and\ \bibinfo {author} {\bibfnamefont {M.}~\bibnamefont
  {Lucente}},\ }\href {\doibase 10.1016/j.nuclphysb.2014.06.003} {\bibfield
  {journal} {\bibinfo  {journal} {Nucl. Phys.}\ }\textbf {\bibinfo {volume}
  {B885}},\ \bibinfo {pages} {651} (\bibinfo {year} {2014})},\ \Eprint
  {http://arxiv.org/abs/1401.1507} {arXiv:1401.1507 [hep-ph]} \BibitemShut
  {NoStop}%
\bibitem [{\citenamefont {Humbert}\ \emph
  {et~al.}(2015{\natexlab{a}})\citenamefont {Humbert}, \citenamefont
  {Lindner},\ and\ \citenamefont {Smirnov}}]{Humbert:2015epa}%
  \BibitemOpen
  \bibfield  {author} {\bibinfo {author} {\bibfnamefont {P.}~\bibnamefont
  {Humbert}}, \bibinfo {author} {\bibfnamefont {M.}~\bibnamefont {Lindner}}, \
  and\ \bibinfo {author} {\bibfnamefont {J.}~\bibnamefont {Smirnov}},\ }\href
  {\doibase 10.1007/JHEP06(2015)035} {\bibfield  {journal} {\bibinfo  {journal}
  {JHEP}\ }\textbf {\bibinfo {volume} {06}},\ \bibinfo {pages} {035} (\bibinfo
  {year} {2015}{\natexlab{a}})},\ \Eprint {http://arxiv.org/abs/1503.03066}
  {arXiv:1503.03066 [hep-ph]} \BibitemShut {NoStop}%
\bibitem [{\citenamefont {Dias}\ \emph {et~al.}(2012)\citenamefont {Dias},
  \citenamefont {de~S.~Pires}, \citenamefont {Rodrigues~da Silva},\ and\
  \citenamefont {Sampieri}}]{Dias:2012xp}%
  \BibitemOpen
  \bibfield  {author} {\bibinfo {author} {\bibfnamefont {A.~G.}\ \bibnamefont
  {Dias}}, \bibinfo {author} {\bibfnamefont {C.~A.}\ \bibnamefont
  {de~S.~Pires}}, \bibinfo {author} {\bibfnamefont {P.~S.}\ \bibnamefont
  {Rodrigues~da Silva}}, \ and\ \bibinfo {author} {\bibfnamefont
  {A.}~\bibnamefont {Sampieri}},\ }\href {\doibase 10.1103/PhysRevD.86.035007}
  {\bibfield  {journal} {\bibinfo  {journal} {Phys. Rev.}\ }\textbf {\bibinfo
  {volume} {D86}},\ \bibinfo {pages} {035007} (\bibinfo {year} {2012})},\
  \Eprint {http://arxiv.org/abs/1206.2590} {arXiv:1206.2590 [hep-ph]}
  \BibitemShut {NoStop}%
\bibitem [{\citenamefont {Garayoa}\ \emph {et~al.}(2007)\citenamefont
  {Garayoa}, \citenamefont {Gonzalez-Garcia},\ and\ \citenamefont
  {Rius}}]{Garayoa:2006xs}%
  \BibitemOpen
  \bibfield  {author} {\bibinfo {author} {\bibfnamefont {J.}~\bibnamefont
  {Garayoa}}, \bibinfo {author} {\bibfnamefont {M.}~\bibnamefont
  {Gonzalez-Garcia}}, \ and\ \bibinfo {author} {\bibfnamefont {N.}~\bibnamefont
  {Rius}},\ }\href {\doibase 10.1088/1126-6708/2007/02/021} {\bibfield
  {journal} {\bibinfo  {journal} {JHEP}\ }\textbf {\bibinfo {volume} {02}},\
  \bibinfo {pages} {021} (\bibinfo {year} {2007})},\ \Eprint
  {http://arxiv.org/abs/hep-ph/0611311} {arXiv:hep-ph/0611311} \BibitemShut
  {NoStop}%
\bibitem [{\citenamefont {Humbert}\ \emph
  {et~al.}(2015{\natexlab{b}})\citenamefont {Humbert}, \citenamefont {Lindner},
  \citenamefont {Patra},\ and\ \citenamefont {Smirnov}}]{Humbert:2015yva}%
  \BibitemOpen
  \bibfield  {author} {\bibinfo {author} {\bibfnamefont {P.}~\bibnamefont
  {Humbert}}, \bibinfo {author} {\bibfnamefont {M.}~\bibnamefont {Lindner}},
  \bibinfo {author} {\bibfnamefont {S.}~\bibnamefont {Patra}}, \ and\ \bibinfo
  {author} {\bibfnamefont {J.}~\bibnamefont {Smirnov}},\ }\href {\doibase
  10.1007/JHEP09(2015)064} {\bibfield  {journal} {\bibinfo  {journal} {JHEP}\
  }\textbf {\bibinfo {volume} {09}},\ \bibinfo {pages} {064} (\bibinfo {year}
  {2015}{\natexlab{b}})},\ \Eprint {http://arxiv.org/abs/1505.07453}
  {arXiv:1505.07453 [hep-ph]} \BibitemShut {NoStop}%
\bibitem [{Note1()}]{Note1}%
  \BibitemOpen
  \bibinfo {note} {Different models with a similar underlying idea can be found
  at~\cite
  {Berezinsky:1993fm,Chulia:2016giq,Gu:2010ys,Aranda:2009yb}}\BibitemShut
  {NoStop}%
\bibitem [{\citenamefont {Geng}\ and\ \citenamefont {Ng}(1988)}]{Geng:1988gr}%
  \BibitemOpen
  \bibfield  {author} {\bibinfo {author} {\bibfnamefont {C.~Q.}\ \bibnamefont
  {Geng}}\ and\ \bibinfo {author} {\bibfnamefont {J.~N.}\ \bibnamefont {Ng}},\
  }\href {\doibase 10.1016/0370-2693(88)90817-9} {\bibfield  {journal}
  {\bibinfo  {journal} {Phys. Lett.}\ }\textbf {\bibinfo {volume} {B211}},\
  \bibinfo {pages} {111} (\bibinfo {year} {1988})}\BibitemShut {NoStop}%
\bibitem [{\citenamefont {Tanabashi}\ \emph {et~al.}(2018)\citenamefont
  {Tanabashi} \emph {et~al.}}]{PhysRevD.98.030001}%
  \BibitemOpen
  \bibfield  {author} {\bibinfo {author} {\bibfnamefont {M.}~\bibnamefont
  {Tanabashi}} \emph {et~al.} (\bibinfo {collaboration} {Particle Data
  Group}),\ }\href {\doibase 10.1103/PhysRevD.98.030001} {\bibfield  {journal}
  {\bibinfo  {journal} {Phys. Rev. D}\ }\textbf {\bibinfo {volume} {98}},\
  \bibinfo {pages} {030001} (\bibinfo {year} {2018})}\BibitemShut {NoStop}%
\bibitem [{\citenamefont {Cheung}\ \emph {et~al.}(2015)\citenamefont {Cheung},
  \citenamefont {Ko}, \citenamefont {Lee},\ and\ \citenamefont
  {Tseng}}]{Cheung:2015dta}%
  \BibitemOpen
  \bibfield  {author} {\bibinfo {author} {\bibfnamefont {K.}~\bibnamefont
  {Cheung}}, \bibinfo {author} {\bibfnamefont {P.}~\bibnamefont {Ko}}, \bibinfo
  {author} {\bibfnamefont {J.~S.}\ \bibnamefont {Lee}}, \ and\ \bibinfo
  {author} {\bibfnamefont {P.-Y.}\ \bibnamefont {Tseng}},\ }\href {\doibase
  10.1007/JHEP10(2015)057} {\bibfield  {journal} {\bibinfo  {journal} {JHEP}\
  }\textbf {\bibinfo {volume} {10}},\ \bibinfo {pages} {057} (\bibinfo {year}
  {2015})},\ \Eprint {http://arxiv.org/abs/1507.06158} {arXiv:1507.06158
  [hep-ph]} \BibitemShut {NoStop}%
\bibitem [{\citenamefont {Akhmedov}\ \emph {et~al.}(1993)\citenamefont
  {Akhmedov}, \citenamefont {Berezhiani}, \citenamefont {Mohapatra},\ and\
  \citenamefont {Senjanovic}}]{Akhmedov:1992hi}%
  \BibitemOpen
  \bibfield  {author} {\bibinfo {author} {\bibfnamefont {E.~K.}\ \bibnamefont
  {Akhmedov}}, \bibinfo {author} {\bibfnamefont {Z.~G.}\ \bibnamefont
  {Berezhiani}}, \bibinfo {author} {\bibfnamefont {R.~N.}\ \bibnamefont
  {Mohapatra}}, \ and\ \bibinfo {author} {\bibfnamefont {G.}~\bibnamefont
  {Senjanovic}},\ }\href {\doibase 10.1016/0370-2693(93)90887-N} {\bibfield
  {journal} {\bibinfo  {journal} {Phys. Lett.}\ }\textbf {\bibinfo {volume}
  {B299}},\ \bibinfo {pages} {90} (\bibinfo {year} {1993})},\ \Eprint
  {http://arxiv.org/abs/hep-ph/9209285} {arXiv:hep-ph/9209285 [hep-ph]}
  \BibitemShut {NoStop}%
\bibitem [{\citenamefont {Drewes}\ \emph {et~al.}(2017)\citenamefont {Drewes}
  \emph {et~al.}}]{Adhikari:2016bei}%
  \BibitemOpen
  \bibfield  {author} {\bibinfo {author} {\bibfnamefont {M.}~\bibnamefont
  {Drewes}} \emph {et~al.},\ }\href {\doibase 10.1088/1475-7516/2017/01/025}
  {\bibfield  {journal} {\bibinfo  {journal} {JCAP}\ }\textbf {\bibinfo
  {volume} {01}},\ \bibinfo {pages} {025} (\bibinfo {year} {2017})},\ \Eprint
  {http://arxiv.org/abs/1602.04816} {arXiv:1602.04816 [hep-ph]} \BibitemShut
  {NoStop}%
\bibitem [{\citenamefont {Lattanzi}\ \emph {et~al.}(2013)\citenamefont
  {Lattanzi}, \citenamefont {Riemer-Sorensen}, \citenamefont {Tortola},\ and\
  \citenamefont {Valle}}]{Lattanzi:2013uza}%
  \BibitemOpen
  \bibfield  {author} {\bibinfo {author} {\bibfnamefont {M.}~\bibnamefont
  {Lattanzi}}, \bibinfo {author} {\bibfnamefont {S.}~\bibnamefont
  {Riemer-Sorensen}}, \bibinfo {author} {\bibfnamefont {M.}~\bibnamefont
  {Tortola}}, \ and\ \bibinfo {author} {\bibfnamefont {J.~W.~F.}\ \bibnamefont
  {Valle}},\ }\href {\doibase 10.1103/PhysRevD.88.063528} {\bibfield  {journal}
  {\bibinfo  {journal} {Phys. Rev.}\ }\textbf {\bibinfo {volume} {D88}},\
  \bibinfo {pages} {063528} (\bibinfo {year} {2013})},\ \Eprint
  {http://arxiv.org/abs/1303.4685} {arXiv:1303.4685 [astro-ph.HE]} \BibitemShut
  {NoStop}%
\bibitem [{\citenamefont {Cirelli}\ \emph {et~al.}(2012)\citenamefont
  {Cirelli}, \citenamefont {Moulin}, \citenamefont {Panci}, \citenamefont
  {Serpico},\ and\ \citenamefont {Viana}}]{Cirelli:2012ut}%
  \BibitemOpen
  \bibfield  {author} {\bibinfo {author} {\bibfnamefont {M.}~\bibnamefont
  {Cirelli}}, \bibinfo {author} {\bibfnamefont {E.}~\bibnamefont {Moulin}},
  \bibinfo {author} {\bibfnamefont {P.}~\bibnamefont {Panci}}, \bibinfo
  {author} {\bibfnamefont {P.~D.}\ \bibnamefont {Serpico}}, \ and\ \bibinfo
  {author} {\bibfnamefont {A.}~\bibnamefont {Viana}},\ }\href {\doibase
  10.1103/PhysRevD.86.083506, 10.1103/PhysRevD.86.109901} {\bibfield  {journal}
  {\bibinfo  {journal} {Phys. Rev.}\ }\textbf {\bibinfo {volume} {D86}},\
  \bibinfo {pages} {083506} (\bibinfo {year} {2012})},\ \Eprint
  {http://arxiv.org/abs/1205.5283} {arXiv:1205.5283 [astro-ph.CO]} \BibitemShut
  {NoStop}%
\bibitem [{\citenamefont {Schechter}\ and\ \citenamefont
  {Valle}(1982)}]{Schechter:1981cv}%
  \BibitemOpen
  \bibfield  {author} {\bibinfo {author} {\bibfnamefont {J.}~\bibnamefont
  {Schechter}}\ and\ \bibinfo {author} {\bibfnamefont {J.~W.~F.}\ \bibnamefont
  {Valle}},\ }\href {\doibase 10.1103/PhysRevD.25.774} {\bibfield  {journal}
  {\bibinfo  {journal} {Phys. Rev.}\ }\textbf {\bibinfo {volume} {D25}},\
  \bibinfo {pages} {774} (\bibinfo {year} {1982})}\BibitemShut {NoStop}%
\bibitem [{Note2()}]{Note2}%
  \BibitemOpen
  \bibinfo {note} {This comes from the fact that in the Majoron's restframe, it
  cannot decay into particles heavier than itself, like $\zeta _3$, $\zeta _4$
  and $\zeta _5$. Hence, only off-shell production of the heavy particles and
  its subsequent decay are allowed. The larger is difference in mass, the
  better this approximation will work.}\BibitemShut {Stop}%
\bibitem [{\citenamefont {Hall}\ \emph {et~al.}(2010)\citenamefont {Hall},
  \citenamefont {Jedamzik}, \citenamefont {March-Russell},\ and\ \citenamefont
  {West}}]{Hall:2009bx}%
  \BibitemOpen
  \bibfield  {author} {\bibinfo {author} {\bibfnamefont {L.~J.}\ \bibnamefont
  {Hall}}, \bibinfo {author} {\bibfnamefont {K.}~\bibnamefont {Jedamzik}},
  \bibinfo {author} {\bibfnamefont {J.}~\bibnamefont {March-Russell}}, \ and\
  \bibinfo {author} {\bibfnamefont {S.~M.}\ \bibnamefont {West}},\ }\href
  {\doibase 10.1007/JHEP03(2010)080} {\bibfield  {journal} {\bibinfo  {journal}
  {JHEP}\ }\textbf {\bibinfo {volume} {03}},\ \bibinfo {pages} {080} (\bibinfo
  {year} {2010})},\ \Eprint {http://arxiv.org/abs/0911.1120} {arXiv:0911.1120
  [hep-ph]} \BibitemShut {NoStop}%
\bibitem [{\citenamefont {Gondolo}\ and\ \citenamefont
  {Gelmini}(1991)}]{Gondolo:1990dk}%
  \BibitemOpen
  \bibfield  {author} {\bibinfo {author} {\bibfnamefont {P.}~\bibnamefont
  {Gondolo}}\ and\ \bibinfo {author} {\bibfnamefont {G.}~\bibnamefont
  {Gelmini}},\ }\href {\doibase 10.1016/0550-3213(91)90438-4} {\bibfield
  {journal} {\bibinfo  {journal} {Nucl. Phys.}\ }\textbf {\bibinfo {volume}
  {B360}},\ \bibinfo {pages} {145} (\bibinfo {year} {1991})}\BibitemShut
  {NoStop}%
\bibitem [{\citenamefont {Heeck}\ and\ \citenamefont
  {Teresi}(2017)}]{Heeck:2017xbu}%
  \BibitemOpen
  \bibfield  {author} {\bibinfo {author} {\bibfnamefont {J.}~\bibnamefont
  {Heeck}}\ and\ \bibinfo {author} {\bibfnamefont {D.}~\bibnamefont {Teresi}},\
  }\href {\doibase 10.1103/PhysRevD.96.035018} {\bibfield  {journal} {\bibinfo
  {journal} {Phys. Rev.}\ }\textbf {\bibinfo {volume} {D96}},\ \bibinfo {pages}
  {035018} (\bibinfo {year} {2017})},\ \Eprint
  {http://arxiv.org/abs/1706.09909} {arXiv:1706.09909 [hep-ph]} \BibitemShut
  {NoStop}%
\bibitem [{\citenamefont {Weinberg}(2013)}]{Weinberg:2013kea}%
  \BibitemOpen
  \bibfield  {author} {\bibinfo {author} {\bibfnamefont {S.}~\bibnamefont
  {Weinberg}},\ }\href {\doibase 10.1103/PhysRevLett.110.241301} {\bibfield
  {journal} {\bibinfo  {journal} {Phys. Rev. Lett.}\ }\textbf {\bibinfo
  {volume} {110}},\ \bibinfo {pages} {241301} (\bibinfo {year} {2013})},\
  \Eprint {http://arxiv.org/abs/1305.1971} {arXiv:1305.1971 [astro-ph.CO]}
  \BibitemShut {NoStop}%
\bibitem [{\citenamefont {Husdal}(2016)}]{Husdal:2016haj}%
  \BibitemOpen
  \bibfield  {author} {\bibinfo {author} {\bibfnamefont {L.}~\bibnamefont
  {Husdal}},\ }\href {\doibase 10.3390/galaxies4040078} {\bibfield  {journal}
  {\bibinfo  {journal} {Galaxies}\ }\textbf {\bibinfo {volume} {4}},\ \bibinfo
  {pages} {78} (\bibinfo {year} {2016})},\ \Eprint
  {http://arxiv.org/abs/1609.04979} {arXiv:1609.04979 [astro-ph.CO]}
  \BibitemShut {NoStop}%
\bibitem [{\citenamefont {Frigerio}\ \emph {et~al.}(2011)\citenamefont
  {Frigerio}, \citenamefont {Hambye},\ and\ \citenamefont
  {Masso}}]{Frigerio:2011in}%
  \BibitemOpen
  \bibfield  {author} {\bibinfo {author} {\bibfnamefont {M.}~\bibnamefont
  {Frigerio}}, \bibinfo {author} {\bibfnamefont {T.}~\bibnamefont {Hambye}}, \
  and\ \bibinfo {author} {\bibfnamefont {E.}~\bibnamefont {Masso}},\ }\href
  {\doibase 10.1103/PhysRevX.1.021026} {\bibfield  {journal} {\bibinfo
  {journal} {Phys. Rev.}\ }\textbf {\bibinfo {volume} {X1}},\ \bibinfo {pages}
  {021026} (\bibinfo {year} {2011})},\ \Eprint {http://arxiv.org/abs/1107.4564}
  {arXiv:1107.4564 [hep-ph]} \BibitemShut {NoStop}%
\bibitem [{\citenamefont {Boulebnane}\ \emph {et~al.}(2018)\citenamefont
  {Boulebnane}, \citenamefont {Heeck}, \citenamefont {Nguyen},\ and\
  \citenamefont {Teresi}}]{Boulebnane:2017fxw}%
  \BibitemOpen
  \bibfield  {author} {\bibinfo {author} {\bibfnamefont {S.}~\bibnamefont
  {Boulebnane}}, \bibinfo {author} {\bibfnamefont {J.}~\bibnamefont {Heeck}},
  \bibinfo {author} {\bibfnamefont {A.}~\bibnamefont {Nguyen}}, \ and\ \bibinfo
  {author} {\bibfnamefont {D.}~\bibnamefont {Teresi}},\ }\href {\doibase
  10.1088/1475-7516/2018/04/006} {\bibfield  {journal} {\bibinfo  {journal}
  {JCAP}\ }\textbf {\bibinfo {volume} {1804}},\ \bibinfo {pages} {006}
  (\bibinfo {year} {2018})},\ \Eprint {http://arxiv.org/abs/1709.07283}
  {arXiv:1709.07283 [hep-ph]} \BibitemShut {NoStop}%
\bibitem [{\citenamefont {Chatrchyan}\ \emph {et~al.}(2014)\citenamefont
  {Chatrchyan} \emph {et~al.}}]{Chatrchyan:2014tja}%
  \BibitemOpen
  \bibfield  {author} {\bibinfo {author} {\bibfnamefont {S.}~\bibnamefont
  {Chatrchyan}} \emph {et~al.} (\bibinfo {collaboration} {CMS}),\ }\href
  {\doibase 10.1140/epjc/s10052-014-2980-6} {\bibfield  {journal} {\bibinfo
  {journal} {Eur. Phys. J.}\ }\textbf {\bibinfo {volume} {C74}},\ \bibinfo
  {pages} {2980} (\bibinfo {year} {2014})},\ \Eprint
  {http://arxiv.org/abs/1404.1344} {arXiv:1404.1344 [hep-ex]} \BibitemShut
  {NoStop}%
\bibitem [{\citenamefont {Baek}\ \emph {et~al.}(2014)\citenamefont {Baek},
  \citenamefont {Ko},\ and\ \citenamefont {Park}}]{Baek:2014jga}%
  \BibitemOpen
  \bibfield  {author} {\bibinfo {author} {\bibfnamefont {S.}~\bibnamefont
  {Baek}}, \bibinfo {author} {\bibfnamefont {P.}~\bibnamefont {Ko}}, \ and\
  \bibinfo {author} {\bibfnamefont {W.-I.}\ \bibnamefont {Park}},\ }\href
  {\doibase 10.1103/PhysRevD.90.055014} {\bibfield  {journal} {\bibinfo
  {journal} {Phys. Rev.}\ }\textbf {\bibinfo {volume} {D90}},\ \bibinfo {pages}
  {055014} (\bibinfo {year} {2014})},\ \Eprint {http://arxiv.org/abs/1405.3530}
  {arXiv:1405.3530 [hep-ph]} \BibitemShut {NoStop}%
\bibitem [{\citenamefont {Staub}(2014)}]{Staub:2013tta}%
  \BibitemOpen
  \bibfield  {author} {\bibinfo {author} {\bibfnamefont {F.}~\bibnamefont
  {Staub}},\ }\href {\doibase 10.1016/j.cpc.2014.02.018} {\bibfield  {journal}
  {\bibinfo  {journal} {Comput. Phys. Commun.}\ }\textbf {\bibinfo {volume}
  {185}},\ \bibinfo {pages} {1773} (\bibinfo {year} {2014})},\ \Eprint
  {http://arxiv.org/abs/1309.7223} {arXiv:1309.7223 [hep-ph]} \BibitemShut
  {NoStop}%
\bibitem [{\citenamefont {Chuliá}\ \emph {et~al.}(2016)\citenamefont
  {Chuliá}, \citenamefont {Srivastava},\ and\ \citenamefont
  {Valle}}]{Chulia:2016giq}%
  \BibitemOpen
  \bibfield  {author} {\bibinfo {author} {\bibfnamefont {S.~C.}\ \bibnamefont
  {Chuliá}}, \bibinfo {author} {\bibfnamefont {R.}~\bibnamefont {Srivastava}},
  \ and\ \bibinfo {author} {\bibfnamefont {J.~W.~F.}\ \bibnamefont {Valle}},\
  }\href {\doibase 10.1016/j.physletb.2016.08.028} {\bibfield  {journal}
  {\bibinfo  {journal} {Phys. Lett.}\ }\textbf {\bibinfo {volume} {B761}},\
  \bibinfo {pages} {431} (\bibinfo {year} {2016})},\ \Eprint
  {http://arxiv.org/abs/1606.06904} {arXiv:1606.06904 [hep-ph]} \BibitemShut
  {NoStop}%
\bibitem [{\citenamefont {Gu}\ \emph {et~al.}(2010)\citenamefont {Gu},
  \citenamefont {Ma},\ and\ \citenamefont {Sarkar}}]{Gu:2010ys}%
  \BibitemOpen
  \bibfield  {author} {\bibinfo {author} {\bibfnamefont {P.-H.}\ \bibnamefont
  {Gu}}, \bibinfo {author} {\bibfnamefont {E.}~\bibnamefont {Ma}}, \ and\
  \bibinfo {author} {\bibfnamefont {U.}~\bibnamefont {Sarkar}},\ }\href
  {\doibase 10.1016/j.physletb.2010.05.012} {\bibfield  {journal} {\bibinfo
  {journal} {Phys. Lett.}\ }\textbf {\bibinfo {volume} {B690}},\ \bibinfo
  {pages} {145} (\bibinfo {year} {2010})},\ \Eprint
  {http://arxiv.org/abs/1004.1919} {arXiv:1004.1919 [hep-ph]} \BibitemShut
  {NoStop}%
\end{thebibliography}%

\end{document}